\documentclass[twocolumn]{aastex631}
\usepackage{xspace}
\usepackage[hyphens]{url}

\pdfoutput=1 
\usepackage[figure,figure*]{hypcap}
\usepackage{footnote}
\usepackage{hyperref}
\usepackage{subfigure}
\usepackage{amsmath}

\usepackage{epsfig}
\usepackage{natbib}

\begin{document}

\def\Msun{\hbox{M$_{\odot}$}}
\def\Lsun{\hbox{L$_{\odot}$}}
\def\kms{km~s$^{\rm -1}$}
\def\hcop{HCO$^{+}$}
\def\n2hp{N$_{2}$H$^{+}$}
\def\micron{$\mu$m\xspace}
\def\13CO{$^{13}$CO}
\def\etamb{$\eta_{\rm mb}$}
\def\Inu{I$_{\nu}$}
\def\kapnu{$\kappa _{\nu}$}
\def\ffore{f$_{\rm{fore}}$}
\def\tastar{T$_{A}^{*}$}
\def\nh3{NH$_{3}$}
\def\deg{$^{\circ}$\xspace}
\def\arcsec{$^{\prime\prime}$\xspace}
\def\arcmin{$^{\prime}$\xspace}
\def\Vlsr{\hbox{V$_{LSR}$}}

\title[3-D CMZ IV: Dust Extinction]{3-D CMZ IV: Distinguishing Near vs. Far Distances in the Galactic Center Using Spitzer and Herschel}

\author[0000-0002-5776-9473]{Dani Lipman}
\affiliation{University of Connecticut, Department of Physics, 196A Hillside Road, Unit 3046
Storrs, CT 06269-3046, USA}

\author[0000-0002-6073-9320]{Cara Battersby}
\affiliation{University of Connecticut, Department of Physics, 196A Hillside Road, Unit 3046
Storrs, CT 06269-3046, USA}
\affiliation{Center for Astrophysics $|$ Harvard \& Smithsonian, MS-78, 60 Garden St., Cambridge, MA 02138 USA}

\author[0000-0001-7330-8856]{Daniel L. Walker}
\affiliation{UK ALMA Regional Centre Node, Jodrell Bank Centre for Astrophysics, Oxford Road, The University of Manchester, Manchester M13 9PL, United Kingdom}
\affiliation{University of Connecticut, Department of Physics, 196A Hillside Road, Unit 3046
Storrs, CT 06269-3046, USA}

\author[0000-0001-6113-6241]{Mattia C.\ Sormani}
\affiliation{Universit{\`a} dell’Insubria, via Valleggio 11, 22100 Como, Italy}

\author[0000-0001-8135-6612]{John Bally}
\affiliation{CASA, University of Colorado, 389-UCB, Boulder, CO 80309}

\author[0000-0003-0410-4504]{Ashley Barnes}
\affiliation{European Southern Observatory (ESO), Karl-Schwarzschild-Straße 2, 85748 Garching, Germany}

\author[0000-0001-6431-9633]{Adam Ginsburg}
\affiliation{University of Florida Department of Astronomy, Bryant Space Science Center, Gainesville, FL, 32611, USA}

\author[0000-0001-6708-1317]{Simon C. O. Glover}
\affiliation{Universit\"{a}t Heidelberg, Zentrum f\"{u}r Astronomie, Institut f\"{u}r Theoretische Astrophysik, Albert-Ueberle-Str.\ 2, 69120 Heidelberg, Germany}

\author[0000-0001-9656-7682]{Jonathan~D.~Henshaw}
\affiliation{Astrophysics Research Institute, Liverpool John Moores University, 146 Brownlow Hill, Liverpool L3 5RF, UK}
\affiliation{Max-Planck-Institut f\"ur Astronomie, K\"onigstuhl 17, D-69117 Heidelberg, Germany}

\author[0000-0003-0946-4365]{H Perry Hatchfield}
\affiliation{Jet Propulsion Laboratory, California Institute of Technology, 4800 Oak Grove Drive, Pasadena, CA, 91109, USA}
\affiliation{University of Connecticut, Department of Physics, 196A Hillside Road, Unit 3046
Storrs, CT 06269-3046, USA}

\author[0000-0003-4140-5138]{Katharina Immer}
\affiliation{European Southern Observatory (ESO), Karl-Schwarzschild-Straße 2, 85748 Garching, Germany}

\author[0000-0002-0560-3172]{Ralf S.\ Klessen}
\affiliation{Universit\"{a}t Heidelberg, Zentrum f\"{u}r Astronomie, Institut f\"{u}r Theoretische Astrophysik, Albert-Ueberle-Str.\ 2, 69120 Heidelberg, Germany}
\affiliation{Universit\"{a}t Heidelberg, Interdisziplin\"{a}res Zentrum f\"{u}r Wissenschaftliches Rechnen, Im Neuenheimer Feld 225, 69120 Heidelberg, Germany}
\affiliation{Center for Astrophysics $|$ Harvard \& Smithsonian, MS-78, 60 Garden St., Cambridge, MA 02138 USA}
\affiliation{Radcliffe Institute for Advanced Studies at Harvard University, 10 Garden Street, Cambridge, MA 02138, USA}

\author[0000-0001-6353-0170]{Steven N.~Longmore}
\affiliation{Astrophysics Research Institute, Liverpool John Moores University, 146 Brownlow Hill, Liverpool L3 5RF, UK}
\affiliation{Cosmic Origins Of Life (COOL) Research DAO, coolresearch.io}

\author[0000-0001-8782-1992]{Elisabeth A.~C.~Mills}
\affiliation{Department of Physics and Astronomy, University of Kansas, 1251 Wescoe Hall Drive, Lawrence, KS 66045, USA}

\author[0000-0002-0820-1814]{Rowan Smith}
\affiliation{Scottish Universities Physics Alliance (SUPA), School of Physics and Astronomy, University of St Andrews, North Haugh, St Andrews KY16 9SS, UK}
\affiliation{Jodrell Bank Centre for Astrophysics, Department of Physics and Astronomy, University of Manchester, Oxford Road, Manchester M13 9PL, UK}

\author[0000-0002-9483-7164]{R. G. Tress}
\affiliation{Institute of Physics, Laboratory for Galaxy Evolution and Spectral Modelling, EPFL, Observatoire de Sauverny, Chemin Pegasi 51, 1290 Versoix, Switzerland}

\author[0009-0005-9578-2192]{Danya Alboslani}
\affiliation{University of Connecticut, Department of Physics, 196A Hillside Road, Unit 3046
Storrs, CT 06269-3046, USA}

\author[0000-0003-2384-6589]{Qizhou Zhang}
\affiliation{Center for Astrophysics $|$ Harvard \& Smithsonian, MS-78, 60 Garden St., Cambridge, MA 02138 USA}

\begin{abstract}
A comprehensive 3-D model of the central 300 pc of the Milky Way, the Central Molecular Zone
(CMZ) is of fundamental importance in understanding energy cycles in galactic nuclei, since the 3-D
structure influences the location and intensity of star formation, feedback, and black hole accretion.
Current observational constraints are insufficient to distinguish between existing 3-D models. Dust
extinction is one diagnostic tool that can help determine the location of dark molecular clouds relative
to the bright Galactic Center emission. By combining Herschel and Spitzer observations, we developed
three new dust extinction techniques to estimate the likely near/far locations for each cloud in the CMZ. We compare our results to four geometric CMZ orbital models. Our extinction methods
show good agreement with each other, and with results from spectral line absorption analysis from Walker et al. (submitted). Our near/far results for CMZ clouds are inconsistent with a projected version of the \citet{Sofue1995} two spiral arms model, and show disagreement in position-velocity space with the \citet{Molinari2011} closed elliptical orbit. Our results are in reasonable agreement with the \citet{Kruijssen2015} open streams. We find that a simplified toy-model elliptical orbit which conserves angular momentum shows promising fits in both position-position and position-velocity space. We conclude that all current CMZ orbital models lack the complexity needed to describe the motion of gas in the CMZ, and further work is needed to construct a complex orbital model to accurately describe gas flows in the CMZ.

\end{abstract}

\keywords{CMZ — Mid Infrared — Galactic Center — Dust Extinction}

\section{Introduction} \label{sec:intro}
  
The central 300pc of the Milky Way, the Central Molecular Zone (CMZ), has extreme physical properties not found elsewhere in our Galaxy, which have been used to study how processes such as star formation vary with environment \citep[see e.g.\ the reviews by][]{Henshaw2023, Bryant2021}. However, while star formation may be expected to vary in different environments, the ability to test these theories is limited by the dynamic range of various properties in our own Galaxy \citep{Rathborne2014} and limited observational resolution when studying other galaxy centers \citep[e.g.][]{Beslic2021, Neumann2023}. These limitations mean the CMZ provides an ideal, nearby laboratory for understanding star and galaxy formation, as we will not be able to study any other galaxy center with the same level of detail in the foreseeable future. High interest in studying the CMZ for its unusual star formation properties and proximity to the central supermassive black hole, SgrA*, has helped produce in-depth pictures of the 2D distribution of gas in the CMZ. However, a self-consistent 3-D model of the CMZ is a key component to address uncertainties in various astrophysics communities.

Star formation in the CMZ is strongly tied to the evolution of its molecular clouds \citep{Sormani&Barnes(2019)}. Observations and simulations both show that material falling into the CMZ can result in collisions, which can stretch the gas, inhibiting star formation \citep{Wallace(2022), gramze2023}, or cause material to collide with molecular gas orbiting the CMZ to act as a trigger for gravitational collapse necessary for star formation \citep{Longmore2013_super_star, Hatchfield2021, Tress2020, Sormani2020}. Without a clear picture of the gas orbits or 3-D distribution of molecular clouds, it is difficult to understand the CMZ's past or future incipient star formation. The 3-D gas distribution and orbital dynamics of material in the CMZ are needed to place our Galaxy in the context of the local Universe. Additionally, a 3-D orbital model is needed to constrain limits on dark matter annihilation and diffuse gamma ray emission from Fermi Telescope measurements \citep[e.g.][]{Ajello(2016),Ackermann(2017),Bartels(2016), Cholis(2015)}, as well as flaring events around SgrA* \citep{Clavel(2013), Ponti(2015)}, making the 3-D geometry of the CMZ integral to answering a plethora of astrophysical questions in addition to star formation histories.

The CMZ is often described as an approximately elliptical ring of gas that orbits the Galactic Center at a radius $\sim 100$~pc (often referred to as the 100~pc stream or 100~pc ring \citep{Kruijssen2015, Molinari2011}). The ring, when projected onto the plane of the sky, can be seen as an infinity-shape in the column density map in Figure \ref{fig:spitz_hers_maps}. Most modern simulations also produce a similar star-forming ring \citep{Armillotta(2019), Salas(2020), Sormani2020}. While there is good agreement about the plane-of-the-sky structure, there is extensive discussion about the top-down, 3-D structure of the gas ring. Current understanding of the 3-D distribution of material in the CMZ has led to various potential geometries to describe the motion of gas that lies on the ring: (i) two inner spiral arms \citep{Sofue1995}, (ii) a closed ellipse \citep{Molinari2011}, (iii) an open stream of orbits (KDL model) \citep{Kruijssen2015}, or (iv) an inner x2 orbit \citep{Binney1991,Tress2020}. Walker et al.\ (submitted) present an example of (iv) using a closed ellipse that aims to add physical and geometric constraints to the \citet{Molinari2011} model, which resulted in a reasonable fit to the gas distribution in both position-position and position-velocity space. 

Overall, the 3-D structure of the orbit on which the gas of the 100~pc stream circles the CMZ is largely unknown. Some constraints have been made as to whether certain molecular clouds exist on the near side of the CMZ (in the foreground of SgrA*) or the far side (in the background). However, distances to most CMZ clouds have yet to be achieved. \citet{Henshaw2016_GasKin_250pc} and \citet{Sofue2022} used kinematic analysis to identify continuous gas streams, but do not find distance estimates for individual clouds. \citet{Yan2017} compare CO emission and OH absorption to investigate the 3-D structure of a handful of individual clouds, and place them on the near/far side of the GC relative to each other. Some studies have found line of sight distances to a few CMZ clouds, with the most constraints placed on the densest CMZ structure, known as the Brick (G0.253+0.016) \citep[][though see \citealp{Zoccali2021}]{Nogueras2021, Martinez-Arranz2022}. Line of sight distances have been estimated to many star-forming regions in the Milky Way’s disk using dust extinction methods \citep{Zhang2022, Kauffmann2008, Flaherty2007, Lombardi2014}. However, gas surface densities in the CMZ tend to be an order of magnitude higher than the solar neighborhood, with typical column densities $\sim 10^{23}$cm$^{-2}$ peaking at $\gtrsim 10^{24}$cm$^{-2}$ in one of the most prolific star forming regions (SgrB2) \citep[Battersby et al.\ submitted]{Henshaw2023, Ginsburg2018}. Thus, for most star-forming regions in the CMZ, the extinction far surpasses $A_{V} \sim 40$~mag. The high column densities mean that near-infrared and mid-infrared (MIR) dust extinction mapping techniques used in the Galactic disk are not easily applicable in the CMZ, as many assumptions fail when applied to the extreme properties of the region.

In this paper, we introduce the application of three new MIR dust extinction mapping methods to determine the relative positions of molecular clouds on the near or far side of the CMZ. This paper builds on the analysis of Battersby et al.\ (submitted; hereafter \textit{Paper I} and \textit{Paper II}) and Walker et al.\ (submitted; hereafter \textit{Paper III}). In \textit{Paper I} and \textit{Paper II} we established a reliable CMZ cloud catalog via derived dust column density maps. In \textit{Paper III} we fit line spectra to identify clear contiguous structures in position-position-velocity space as molecular clouds, and estimate near/far positions of individual clouds using molecular line data. In Section \ref{sec:data} we discuss the archival data used for this analysis, as well as summarize data products from \textit{Papers I} and \textit{II} that are relevant for this work. In Section \ref{sec:new_ext_methods} we introduce two new 8\micron dust extinction methods. Section \ref{sec:70um_ext_method} introduces a new modified dust extinction method to calculate extinction column densities for CMZ clouds. We present results in Section \ref{sec: Results}. In Section \ref{sec: discussion} we compare the results of the three methods presented in this paper (\ref{sec:discuss_NF}), and combine with absorption analysis from \textit{Paper III}, in order to present a qualitative (\ref{sec:discuss_3Dmodels_qualitative}) and quantitative (\ref{sec:discuss_3Dmodels_quantitative}) likely nearside or farside position for all CMZ clouds in the catalog. We present our summary and conclusions in Section \ref{sec:summary_conclusions}.

\section{Data}\label{sec:data}

\subsection{Herschel and Spitzer Data}
The catalog of CMZ clouds was created using observations from the Herschel Infrared Galactic Plane (HiGAL) Survey \citep{Molinari2010, Molinari2016}, which covers the Galactic Plane within $|\ell|<70$\deg and $|b|<1$\deg at 70, 160, 250, 350, and 500 \micron with the Spectral and Photometric Imaging Receiver \citep[SPIRE;][]{SPIRE_Griffin(2010)} and the Photodetector Array Camera and Spectrometer \citep[PACS;][]{PACS_Poglitsch(2010)}. The corresponding beam sizes in each band are $\sim$ 6\arcsec, 12\arcsec , 18\arcsec , 25\arcsec , and 36\arcsec, respectively \citep{Molinari2016}.
In addition to HiGAL observations, we make use of archival 8\micron data from the Spitzer space telescope as part of the GLIMPSE survey \citep{Benjamin2003, Churchwell2009}. We use GLIMPSE II 8\micron residual mosaic images (i.e. images with sources subtracted) covering $|\ell|<2$\deg and $|b|<2$\deg, as well as a field covering $2$\deg$<|\ell|<5$\deg and $|b|<1.5$\deg in order to cover a similar field of view to the HiGAL maps. The PSF FWHM of the 8\micron beam is $\sim 2$\arcsec\footnote{\protect\url{https://irsa.ipac.caltech.edu/data/SPITZER/docs/irac/iracinstrumenthandbook/5/}}. 
GLIMPSE II residual images were produced using the point source extraction program DAOPHOT, which extracts sources down to 2 sigma above the background and removes them from the image\footnote{GLIMPSE II data release: \protect\url{http://ftp.astro.wisc.edu/glimpse/glimpse2_dataprod_v2.1.pdf}}. The background, in the case of the source removal, is determined locally between the PSF fitting radius and an outer sky radius, and is continually redetermined for each iterative step of the source removal process. For further details, we refer the reader to the GLIMPSE II data release and description of the Point Source Photometry provided by the GLIMPSE team\footnote{\protect\url{https://irsa.ipac.caltech.edu/data/SPITZER/GLIMPSE/doc/glimpse_photometry_v1.0.pdf}}.The resulting residual images can be treated as images of the smooth light in the 8\micron map not associated with point sources. The two Spitzer 8\micron residual fields noted here were combined into a composite mosaic with the package \texttt{MontagePy}\footnote{\protect\url{http://montage.ipac.caltech.edu/docs/montagePy-UG.html}}.

\subsection{Derived Column Densities and Cloud Catalog}
Herschel HiGAL data and Spitzer GLIMPSE data have been used together to measure dust temperatures and column densities of both warm and cold dust components within the CMZ \citep[\textit{Paper I},][]{ Barnes2017, Tang2021}. In particular, for this analysis, \textit{Paper I} performed a modified blackbody fit on the observed HiGAL data to produce integrated column density maps in order to determine the spectral energy distribution throughout the CMZ. The resulting fits were used to obtain a column density map of the CMZ, shown in the bottom panel of Figure \ref{fig:spitz_hers_maps}, and cold dust temperatures. 

The analysis in \textit{Paper II} produced a catalog of structures in the column density map employing the \texttt{astrodendro}\footnote{\href{http://www.dendrograms.org/}{http://www.dendrograms.org/}} Python package. The package utilizes a dendrogram algorithm to determine hierarchical structure, where data structures are treated as ``trees" and substructures as ``branches". The local maxima of the highest level substructure are considered ``leaves". In \textit{Paper III}, additional dendrograms were run in separate quadrants of the CMZ to capture lower density clouds and ensure sources were coherent in velocity space. The resulting leaves from \textit{Paper III} are shown by the contours overlying the maps in Figure \ref{fig:spitz_hers_maps}. The 31 leaves are the CMZ clouds used for the analysis presented here, and are listed in Table~\ref{tab: flux_calc_tab}. 

Both physical and kinematic properties were reported for all 31 clouds in the catalog presented in \textit{Paper III}. Individual masks corresponding to the dendrogram leaves were also produced for each cloud. The masks were then applied to GBT data to measure the radio continuum emission towards each cloud. MOPRA 3 mm \citep{Jones2012}, APEX 1 mm \citep{Ginsburg2016}, and ATCA \citep{Ginsburg2015} line survey data were used to extract averaged spectra, integrated intensity maps, velocity maps, and velocity dispersion maps for eight lines for all clouds. In particular, HNCO and H$_{2}$CO were used as kinematic tracers, and a multiple Gaussian fitting was performed for each cloud. There are 12 clouds in the catalog with distinct multiple velocity components in the mean HNCO spectra that are not seen in the H$_{2}$CO. (see e.g. \citet{Henshaw2016_GasKin_250pc} for full spectral decomposition). In these cases, clouds have been divided into separate spatial components based on the projected morphology for the peak intensity of the velocity components (see \textit{Paper III} for more details). We note that the HiGAL density range is limited by the density cut-offs used for the dendrogram analysis. The dendrogram threshold used for most clouds in the catalog is set to 2$\times10^{22}$cm$^{-2}$. The Brick is the lone exception, and uses a threshold of 7$\times10^{22}$cm$^{-2}$ in order to isolate the known kinematic extent of the cloud. 

The dendrogram leaf ID numbers, central ($\mathrm{\ell}$, b) coordinates, and central velocity from the HNCO fits from \textit{Paper III} are summarized in the first four columns of Table~\ref{tab: flux_calc_tab}. Clouds with separate velocity components are noted by alphabetical indices.

\section{New dust extinction methods for application in the CMZ} \label{sec:new_ext_methods}

The bright Galactic background of the CMZ in 8\micron should make it possible to determine the relative position of clouds on the near or far side of the Galactic Center. Figure \ref{fig:spitz_hers_maps} shows a comparison of dust extinction seen in Spitzer 8\micron and Herschel 70\micron, as well as dust emission from the integrated HiGAL column density map for clouds in the CMZ. The Spitzer 8\micron and Herschel 70\micron images reveal dark clouds which absorb and extinguish the bright emission from behind, such as seen with the clearly defined Brick cloud (G0.253+0.016). Conversely, clouds that show up clearly in the Herschel column density map, such as G0.342-0.085 (hereafter, the Sailfish), seem to appear behind the bright continuum emission, as they do not prominently block light in either 8\micron or 70\micron. Ideally, it is possible to compare emission and effective extinction column densities for each cloud to constrain its near/far position. However, the high density of the CMZ presents unique barriers for typical MIR extinction techniques.

In this section, we present a brief explanation of optical depth limitations for MIR techniques in the CMZ (\ref{sec: dustext_optical_depth}). We then present our new dust extinction methods designed to address these limitations, including a flux difference and flux ratio utilizing the Spitzer 8\micron map (\ref{sec:flux_methods}) and a calculated 70\micron extinction column density (\ref{sec:70um_ext_method}).

\subsection{Column density and optical depth limitations in the CMZ}\label{sec: dustext_optical_depth}

Our initial analysis revealed that typical MIR extinction methods are not necessarily applicable to the extremely high column density gas in the CMZ. Therefore, we explored and identified new techniques for determining near vs far positions of molecular clouds based on comparison of the observed emission and absorption in the CMZ. 

Spitzer 8\micron images have been widely used to obtain extinction column density maps using simple foreground and background models \citep{Battersby2010, Ragan2009, Butler&Tan2009}. For IRDCs in the disk, the extinction and emission column densities are seen to map close to a 1:1 correlation \citep{Battersby2010, Battersby2014}. Additionally, similar techniques have been used to help determine probabilistic near vs.\ far kinematic distances of sources using 8\micron extinction features in the Galactic plane \citep{Ellsworth-Bowers2013}. These methods assume the extinction and emission column densities should correlate well, and use this relationship to probabilistically solve for a line of sight distance. Due to the assumption of a good correlation between extinction and emission, this technique works best for non-opaque sources. However, column densities in the CMZ are typically around $\sim 10^{23}$~cm$^{-2}$, corresponding to A$_{V}$ magnitudes $\sim 50$~mag. This means molecular clouds in the CMZ region appear almost entirely opaque at 8\micron, with a corresponding optical depth of $\tau_{8\mu\text{m} } \sim 5.47$. The opacity at 8\micron makes it difficult to determine significant extinction-emission correlations within the CMZ (see Appendix \ref{Appendix_8vs70} for detailed discussion). 

\begin{figure*}
    \centering
    \includegraphics[width=1\textwidth]{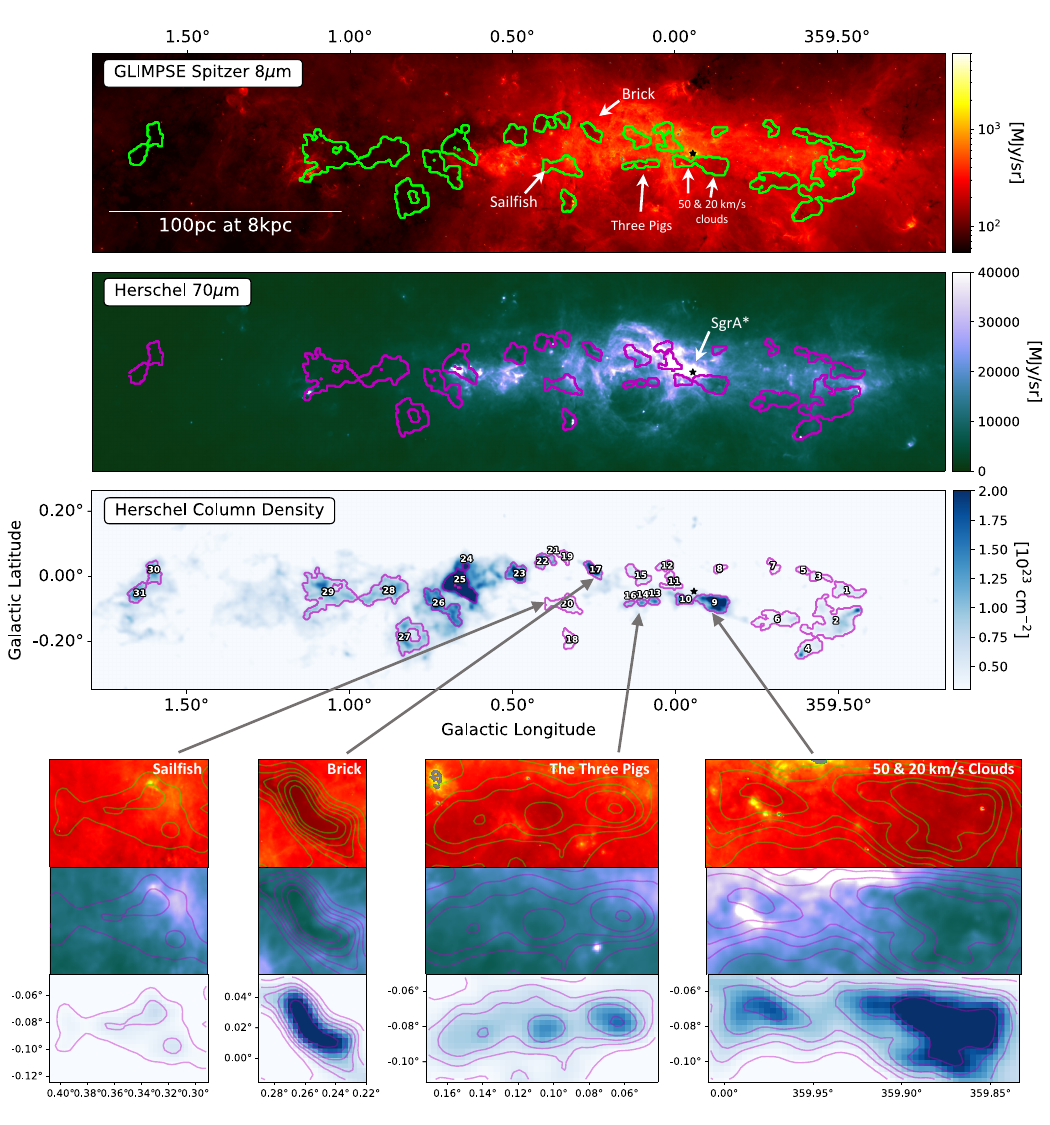}
    \caption{Dense molecular clouds in front of the bright galactic emission appear as dark silhouettes in Spitzer 8\micron and Herschel 70\micron. However, not all clouds of equal column density are equally dark. The top three panels of this figure show comparisons of 1) Spitzer 8\micron , 2) Herschel 70\micron, and 3) derived HiGAL column density with labeled leaf IDs from the dendrogram catalog (bottom). The 8\micron map traces bright PAH emission as well as extinction of emission by dense CMZ clouds. Dense molecular clouds can be seen as dark features in the Herschel column density map. The green and magenta contours indicate individual molecular clouds derived from a dendrogram analysis. The fourth row of panels show zoomed-in cutouts of select clouds: (a) the Sailfish, (b), the Brick, (c) the three pigs (i.e. straw, sticks, and stone clouds), and (d) the 50 km/s and 20 km/s clouds. The Herschel column density contours of the central panels are N(H$_{2}$)$=3$, $5$, $7$, $10$, $15$, $20 \times 10^{22}$ cm$^{-2}$}
    \label{fig:spitz_hers_maps}
\end{figure*}

We can lessen the impact of extreme opacity by studying longer, more optically thin wavelengths. The zoom panels on the bottom row of Figure \ref{fig:spitz_hers_maps} show a comparison of the Spitzer 8\micron (top) and HiGAL 70\micron (middle) maps for various clouds. Unlike the very opaque 8\micron, the 70\micron shows more diffuse emission, with $\tau_{70\mu\text{m} } \sim 0.81$


With both the optical depth limitations and clear evidence of extinction in mind, we chose to explore both the 8\micron and 70\micron data to create new MIR dust extinction techniques in the CMZ. We utilize the Spitzer 8\micron to quantitatively compare the relative darkness of clouds to a modeled background, and use the 70\micron to explore a modified version of typical MIR techniques.

\subsection{8\texorpdfstring{$\mu$}{u}m\xspace extinction methods: the Flux Difference and Flux Ratio} \label{sec:flux_methods}

We present two new, simplistic 8\micron dust extinction methods for determining the likely near vs.\ far positions for all CMZ clouds by comparing the relative emission and absorption of clouds via: 1) a flux difference and 2) a flux ratio. Both methods make use of the GLIMPSE Spitzer 8\micron residual map to compare the observed emission against the estimated foreground and background. 

To estimate the 8\micron emission in the absence of CMZ clouds, we create a smoothed CMZ model using the GLIMPSE Spitzer 8\micron residual mosaic. For each cloud in the catalog, an individual mask was produced corresponding to the area of the dendrogram leaf (See \textit{Paper III}). The clouds were masked out, and the image was then smoothed with a Fast Fourier Transform convolution using \texttt{scipy.signal.fftconvolve} to create a smoothed background map. We use a Gaussian kernel with effective circular radius of 3\arcmin, which returns the best smoothed map while maintaining the general shape of the bright emission. The circular radius can be converted to an effective resolution, $\sigma_{\rm eff}$, to achieve a resolution given by $\sigma =\sqrt{\sigma_{\rm eff}^{2} -\sigma_{\rm beam}^{2}}$, which results in a smoothing kernel much larger than the Spitzer 8\micron resolution ($ \sigma < 1$\arcsec). Finally, the cloud masks are regridded to match the Spitzer 8\micron pixel scaling and used to make cutouts of the 8\micron residual map (foreground and cloud emission) and the smoothed background map (composite of foreground and background emission). 


Our 8\micron methods are similar to each other in that they are both designed to quantitatively measure the darkness of clouds, similar to by-eye determinations of extreme dark clouds, such as the Brick, being confidently declared to sit in the foreground. 

We discuss the individual dust extinction methods below.  

\begin{figure*}
    \centering
    \includegraphics[width=\textwidth]{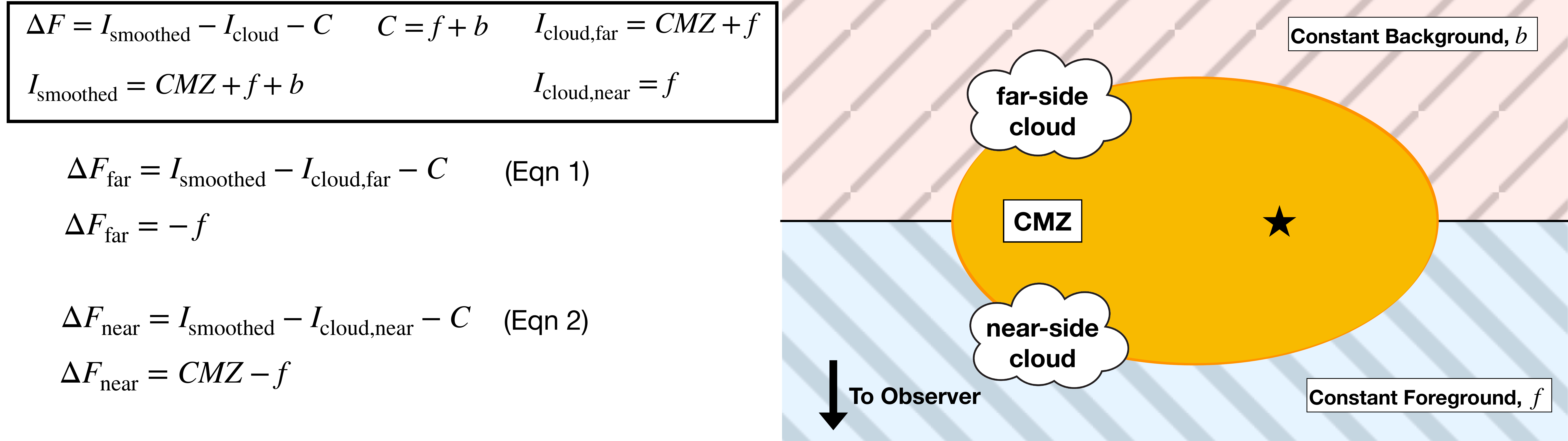}
    \caption{Top-down schematic cartoon to visualize the flux difference and flux ratio methods presented in Section \ref{sec:new_ext_methods}. For both methods, we assume constant foreground (blue) and constant background (pink) distributions with respect to the variable CMZ. The orange area corresponds to the CMZ. For a given cloud, the median estimated 8\micron smoothed model flux, $I_{\rm smoothed}$, corresponds to the combination of the foreground (f), background (b), and the orange ``cmz''. White clouds correspond to any given far-side and near-side clouds. For both the flux difference and flux ratio methods, we assume clouds in the CMZ completely absorb all background radiation at 8\micron.} 
    \label{fig:flux_diff_schematic}

\end{figure*}

\textit{Flux Difference:} We assume both a constant background and constant foreground, implying all observed variability comes from the CMZ area. See Section \ref{sec: uncertainty_assumptions_methods} for discussion of uncertainties and assumptions. Figure \ref{fig:flux_diff_schematic} shows a cartoon top-down schematic of the assumed parameters used for the flux difference and flux ratio methods. The variable flux along the line of sight in 8\micron can almost entirely be attributed to the CMZ itself. Given the high column densities, and optical thickness of all clouds in the CMZ at 8\micron, we assume the clouds absorb all background emission coming from behind the CMZ. The flux measured for a cloud on the far side of the CMZ can be estimated as a combination of the constant foreground and the variable CMZ emission up to the position of the cloud. Similarly, emission of a cloud on the nearest side of the CMZ would be measured as the total constant foreground along the line of sight to the cloud. The smoothed CMZ model obtained from the FFT convolution ($I_{\rm smoothed}$) acts as an estimate of the background, foreground, and variable CMZ together. 
 
  \begin{figure*}
    \centering
    \includegraphics[width=\textwidth]{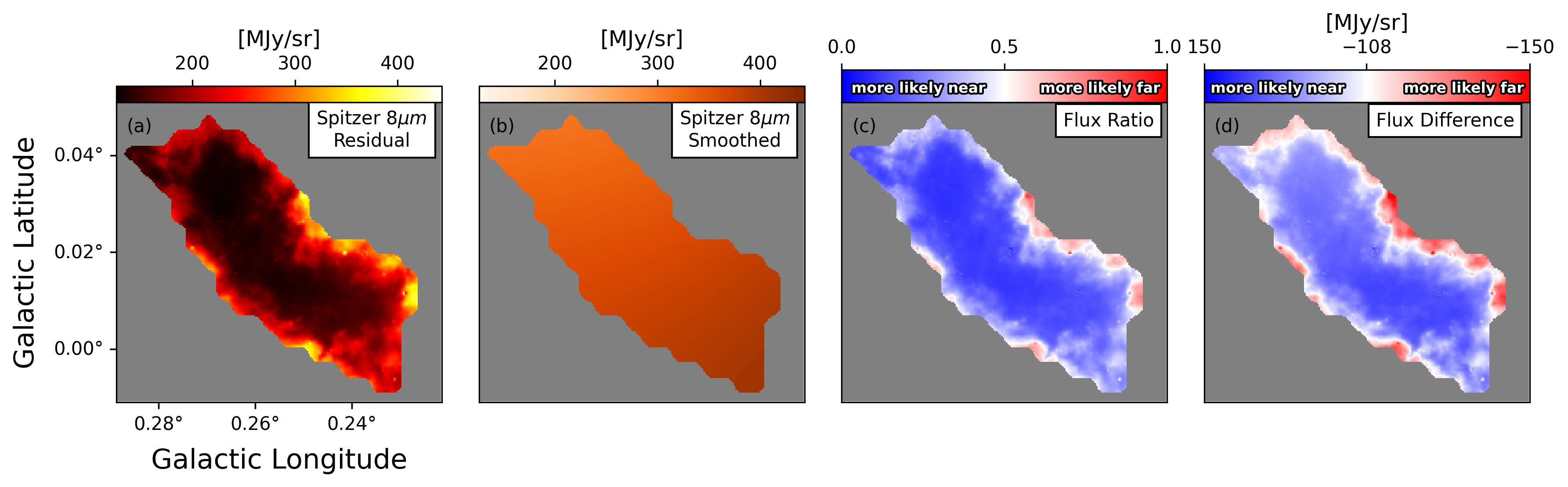}
    \caption{We present two new 8\micron extinction methods which make use of the Spitzer 8\micron residual image and a smoothed background to quantitatively measure the ``darkness" of CMZ clouds compared to the modeled background emission. Here we show an example of the methods applied to the Brick (G0.255+0.02; ID 17). The left two panels show a cutout mask of the cloud for (a) the Spitzer 8\micron residual map, (b) the smoothed background. The right two panels show the resulting (c) flux ratio and (d) flux difference maps for the cloud. Blue colors in panels (c) and (d) represent pixels from the 8\micron image that are significantly darker than the smoothed background (i.e. resulting in more extinction), whereas red colors represent 8\micron pixels which are bright compared to the smoothed background.} 
    \label{fig:brick_flux_methods_4_panel}

\end{figure*}

To estimate the actual flux of each cloud, we take a difference of the median flux value from the 8\micron smoothed model, $I_{\rm smoothed}$, and the median flux of the cloud from the residual Spitzer map, $I_{\rm cloud}$. To determine a value at which the flux difference indicates enough extinction to place clouds on the near side of the CMZ, we also subtract an estimate of the total emission outside of the CMZ, which can be defined as the sum of the constant foreground and background (C=f+b).
 
As a result of the clouds absorbing all background emission at 8\micron, any emission is interpreted to be either from the foreground along the line of sight to the clouds (for near side clouds), or a combination of the foreground and the variable CMZ (for far side clouds). In other words, if a cloud is positioned on the far side of the CMZ, the following equation holds: 
\begin{equation}\label{eqn: I_far} 
    \Delta F_{\rm far} = I_{\rm smoothed} - I_{\rm cloud} - \mathrm{C} = -f.
\end{equation} 
However, if the cloud is positioned on the near side, Equation \ref{eqn: I_near} holds instead:
\begin{equation}\label{eqn: I_near} 
    \Delta F_{\rm near} = I_{\rm smoothed} - I_{\rm cloud} - \mathrm{C} = \mathrm{CMZ} -f
\end{equation}
Here, $f$ is the average foreground, which we estimate as the average of the minimum 8\micron fluxes of the Brick and Sgr B2 that appear in dense areas of the clouds. The minimum values were obtained from an area of high column density and low 8\micron emission in each cloud in the residual image. The median value of the dark regions were taken as the minima, and used to find an average foreground, $f \sim$ 96 MJy/sr. We obtain an estimate for C using the 8\micron residual map by averaging the median flux of several optically thin regions (of 3\arcmin radius) outside of the CMZ, but still within the plane of the GC. These correspond to ($\ell$,$b$) coordinates of (359.42\deg, 0.15\deg), (358.61\deg, 0.01\deg), (358.62\deg, -0.13\deg), and (359.25\deg, -0.32\deg). The areas closest latitudinally to the plane are consistent around 120 -- 130 MJy/sr. We measure an average C=f+b in the residual image of C $\sim$ 125 MJy/sr. 

The values obtained for each cloud using the flux difference method can be compared to the average of Equations~\ref{eqn: I_far} and ~\ref{eqn: I_near}:

\begin{equation}
    \Delta F_{\rm center} = \frac{CMZ}{2} - f = -108.61 ~\mathrm{MJy/sr}
\end{equation}
The average of the near and far limiting cases, $\Delta F_{\rm center}$, can be used to determine likely near/far positions. Assuming negligible emission from the cloud itself, clouds with median flux difference values less than $\Delta F_{\rm center} $ have higher than average foreground emission, and are more likely on the far side. Likewise, clouds with median flux differences greater than $\Delta F_{\rm center} $ may be on the nearer side of the CMZ.



\textit{Flux Ratio:} The second new 8\micron method we present uses the flux ratio between the observed 8\micron intensity and the smoothed background. This method is similar to the flux difference, and thus holds some of the same assumptions and uncertainties. 

The extreme brightness of the CMZ at 8\micron makes it possible to identify areas of strong absorption compared to the average background emission (e.g.\ the Brick), as well as areas of higher emission compared to the background. Taking into account the average foreground in 8\micron, a reasonably simple ratio of the median 8\micron emission for a cloud, $I_{\rm cloud}$, compared to the median modeled background within the cloud mask, $I_{\rm smoothed}$, can easily quantify by how much each cloud deviates above or below the average continuum emission:

\begin{equation}
\text{flux ratio}= \frac{I_{\rm cloud} - f}{I_{\rm smoothed}}  
\end{equation}

Similar to the flux difference method, the average foreground, $f$, is estimated as the average of the minimum 8\micron fluxes of the Brick and Sgr B2 clouds. The flux ratio values alone provide a measure of how bright or dark an individual cloud is compared to the model emission. Calculating the flux ratio for all clouds in the CMZ also allows us to make comparisons between sources to determine a likely near/far position. We determine that clouds with median flux ratios less than 0.5 likely lay on the near side, as they block more of the estimated emission. On the other hand, a cloud with a median flux ratio greater than 0.5 implies emission greater than that of the average foreground, placing the clouds further into in the CMZ. Figure \ref{fig:brick_flux_methods_4_panel} shows an example of the flux difference and flux ratio methods applied to the Brick.


\begin{figure*}
    \centering
    \includegraphics[width=\textwidth]{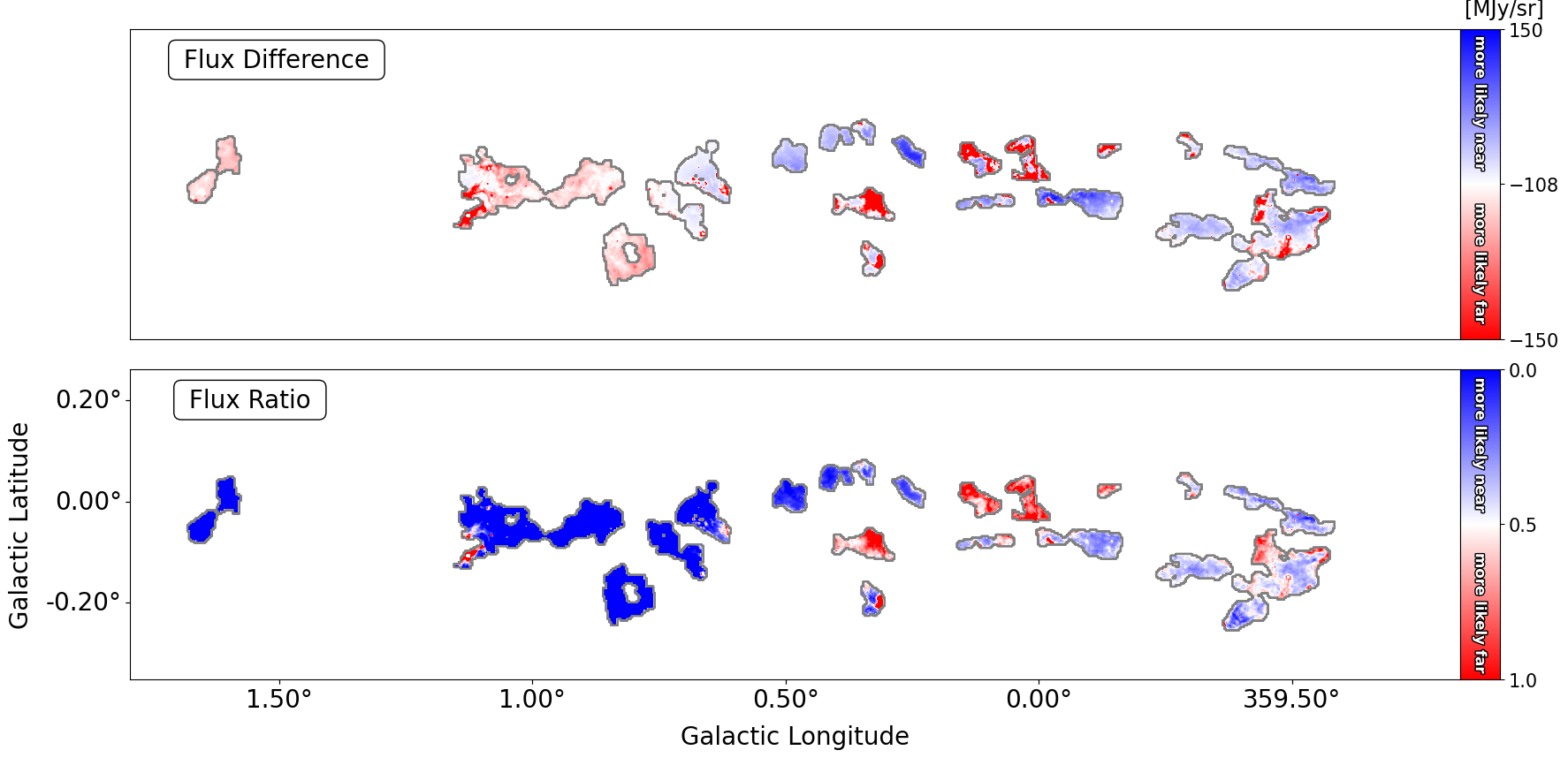}
    \caption{The flux difference and flux ratio methods show good agreement, with the most disagreement in near/far distinctions occurring at longitudes $\ell \gtrsim 0.7$\deg. This figure shows maps of the flux difference (top) and flux ratio (bottom) methods applied to the Spitzer 8\micron data using the dendrogram leaf masks. Values indicative of near-side clouds are blue. Far-side values are red.} 
    \label{fig:flux_methods_full_maps}
\end{figure*}

Figure \ref{fig:flux_methods_full_maps} shows maps of the flux difference (top) and flux ratio (bottom) methods applied to the dendrogram leaf masks for each cloud. The two methods largely agree on near/far distinctions of clouds on the CMZ ring, but disagree for clouds at longitudes of $\ell \gtrsim 0.7$\deg. The methods can be used to infer the structure of material in the CMZ ring. However, the methods cannot be used to infer absolute column densities due to the saturation and opacity in 8\micron (See \ref{Appendix_8vs70}). We report the median value of the flux difference and flux ratio method for each mask (or sub-mask for clouds with multiple velocity components) in Table \ref{tab: flux_calc_tab}.

\subsection{New 70\texorpdfstring{$\mu$}{u}m\xspace extinction method}\label{sec:70um_ext_method}

High column densities and opacities in the CMZ mean that normal NIR and MIR extinction mapping techniques cannot be applied to obtain distance estimates for CMZ clouds in our catalogs (See Appendix \ref{Appendix_8vs70}). Here, we modify a calculated 8\micron extinction method often used to compare extinction and emission of infrared dark clouds in the Galactic disk, and apply the modified method to the 70\micron Herschel map.

MIR dust extinction is usually calculated by estimating the Galactic background and foreground emission, finding the optical depth needed to produce the observed extincted intensity (assuming a given dust opacity), and finally deriving a gas mass surface density, which we can then convert to an equivalent extinction column density for comparison with observed emission \citep[e.g. see][]{Butler&Tan2009, Battersby2010}. For example, \citet{Battersby2010} compare Bolocam 1.1 mm dust emission and Spitzer 8\micron data for IRDCs in the disk of the Galaxy. We explore the same methodology to calculate extinction-based surface densities of clouds at 70\micron by modifying Equation 7 from \citet{Battersby2010} to calculate an extinction-based surface density ($\Sigma$):

\begin{equation} \label{eqn:ExtSig_nu}
    \Sigma_{\nu} = -\frac{1}{\kappa_{\nu}} \ln \left[\frac{\mathrm{I}_{\nu 1, \rm obs} - f_{\rm fore}\mathrm{I}_{\nu 0, \rm obs}}{(1-f_{\rm fore})\mathrm{I}_{\nu 0,\rm obs}}\right]
\end{equation} 

\noindent where $\kappa_{\nu}$ is the dust opacity. The resulting mass surface density of the cloud, $\Sigma$, is dependent on the measured intensity in front of the cloud (I$_{\nu 1,\rm obs}$), the composite intensity of the foreground and estimated background at the location of the cloud (I$_{\nu 0,\rm obs}$), and the foreground intensity ratio (f$_{\rm fore}$), defined as the ratio of intensities out to 8 kpc and out to 16 kpc (i.e.\ twice the distance to the CMZ, assuming a distance to SgrA* of $\sim$ 8 kpc). The original equation assumes a scattering coefficient, $s$, a correction for the Spitzer IRAC array. For our 70\micron analysis we adopt a scattering coefficient $\mathrm{s}=0$. We estimate $\kappa_{\nu} = 1.74$ cm$^{2}$g$^{-1}$ using models from \citet{Ossenkopf&Henning(2004)} of thin ice mantles that have undergone coagulation for $10^{5}$ years at a density of $n_{\rm H2} \sim  10^{6} \, {\rm cm}^{-3}$. 

For this analysis, we assume that the bright 70\micron emission seen in Figure \ref{fig:spitz_hers_maps} is from the CMZ, and continue to treat the 70\micron opacity as an extinction opacity. We note that while the emission in 70\micron is due to the hot component of the gas, which the dense CMZ clouds extinguish, the clouds in 70\micron are not completely optically thick. Thus, it is possible that some of that emission may be due to the clouds themselves. The temperatures of CMZ clouds are typically $\sim$20 -- 25K (See  \textit{Paper I}), while the blackbody emission for 70\micron peaks at $\sim$40K. The method we present is focused on identifying the correlation of the absorption to the cold dust. So while the emission from the clouds is an important uncertainty, particularly for brighter pixels, it should not dominate the observed flux from the dark clouds.  

We implement the new 70\micron method and test its viability in the CMZ by assuming that half of the observed intensity is due to foreground emission, i.e.\ $f_{\rm fore} =0.5$. We make this assumption as a test, as all cataloged clouds exist within a central few hundred parsec radius of the Galactic Center, which should avoid the need to correct for large scale dust extinction variations \citep[e.g. see][]{Schultheis2014}. Thus, the differences in f$_{\rm fore}$ between clouds should be small, and likely within a range from $0.45 <$ $f_{\rm fore} < 0.55$ for most CMZ clouds.

The HiGAL 70\micron map has many bright sources that significantly impact the $\Sigma_{70}$ calculation. We therefore mask out the compact sources found in the 70\micron source catalog from \citet{Molinari2016}\footnote{\protect\url{https://cdsarc.cds.unistra.fr/ftp/J/A+A/591/A149/}}. The resulting source-removed 70\micron map is used for the extinction calculation procedure. For more details, and example extinction calculations using other values for the foreground intensity ratios, see Appendix~\ref{Appendix_8vs70}.

To estimate the smoothed modeled background, each cloud mask is regridded to match the Herschel 70\micron pixel scaling, and used to create cutouts of each cloud from the 70\micron image. The image is then smoothed with the same 3\arcmin radius kernel, as done for the 8\micron smoothed map, to achieve a modeled background map. The regridded cloud masks are then used to make cutouts of the 70\micron source-removed map (I$_{\nu 1,\rm obs}$) and the modeled background map (I$_{\nu 0,\rm obs}$). 

\begin{figure*}
    \includegraphics[width=\textwidth]{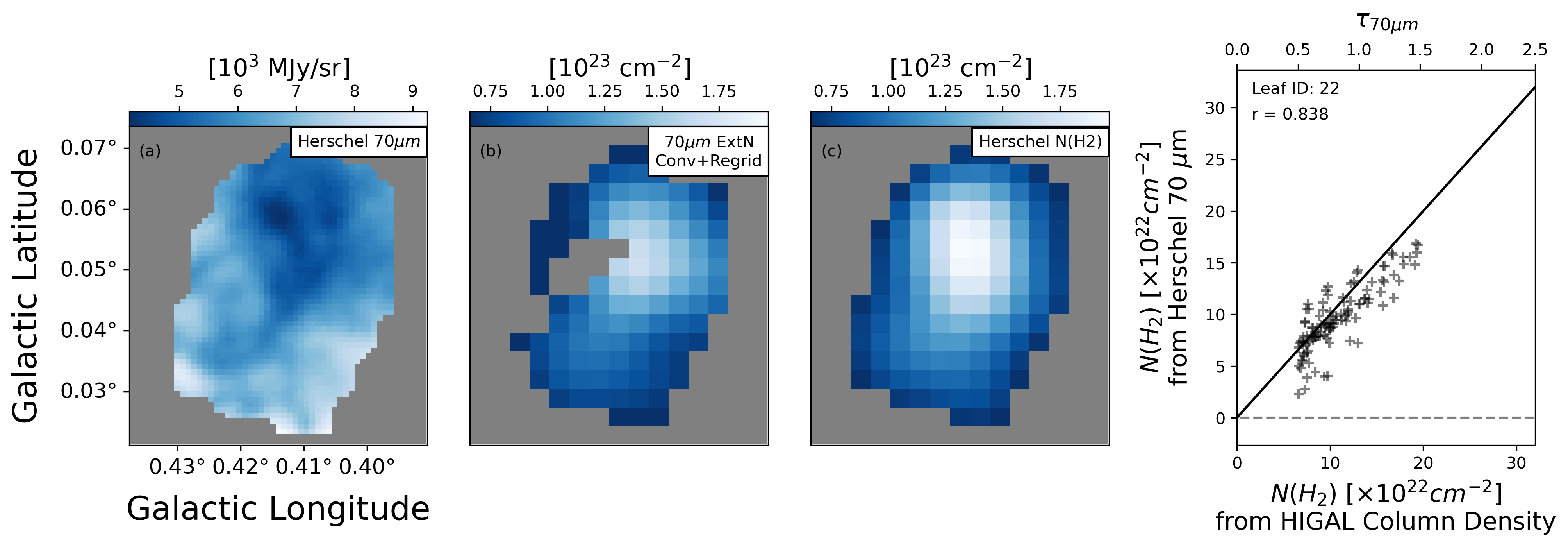}
    \caption{The Herschel 70\micron is less optically thick than the 8\micron, making it a better wavelength to use when applying the extinction mapping approach to CMZ clouds. Here we show examples of the 70\micron extinction method for cloud ID 22: G0.413+0.048. The left three panels show a cutout mask of the clouds for (a) the Herschel 70\micron map, (b) the final calculated 70\micron extinction column density $f_{\rm fore} =0.5$, after which it is convolved to the HiGAL column density FWHM and regridded to match the HiGAL pixel scaling, and (c) the HiGAL column density map. Cataloged 70\micron sources from \citep{Molinari2011} were removed before calculating the extinction column density map in panel (b). Grey pixels represent NaN values. The pixel-by-pixel correlation between the calculated extinction column density and HiGAL emission is shown in the rightmost panel. Corresponding 70\micron optical depths are reported on the top axis. We report the Pearson correlation coefficient (r) between the extinction and emission, where r$=1$ indicates a perfect correlation.} The example cloud, G0.413+0.048, shows strong correlation, which we take to imply a strong likelihood of a position on the nearside of the CMZ.
    \label{fig:70um_method_ID22}
\end{figure*}

Next, we calculate $\Sigma_{70}$ at a given f$_{\rm fore}$ for each pixel using Equation~\ref{eqn:ExtSig_nu}. The $\Sigma$ values are converted to a 70\micron extinction column density ($N(\rm H_{2})_{70\mu m}$) using a simple relation:
\begin{equation}\label{NH2}
    N({\rm H}_{2})_{70\mu m} = \Sigma/(\mu_{\rm H2} m_{\rm H})
\end{equation}
where $\mu_{\rm H2}$ is the mean mass per hydrogen molecule in atomic mass units, taken to be 2.8, and $m_{\rm H}$ is the mass of a hydrogen atom. The result of these calculations produces an extinction column density map for each cloud mask in the catalog. The $N(\rm H_{2})_{70\mu m}$ map is then smoothed to match the 36\arcsec FWHM of the HiGAL integrated column density map to produce an ``observed" comparison to the HiGAL column density maps for each cloud. Figure \ref{fig:70um_method_ID22} shows an example of the 70\micron method applied to Cloud ID 22 (G0.413+0.048) using $f_{\rm fore}=0.5$. The pixel-by-pixel correlation between the calculated extinction column density and HiGAL emission is shown in the rightmost panel. 

To determine reasonable pixel-by-pixel correlations between the extinction and emission, we calculate a single Pearson correlation coefficient, r, between the calculated 70\micron extinction column density and the HiGAL emission column density for each cloud. The Pearson correlation coefficient is defined as the the covariance of two normally distributed variables divided by the product of their standard deviations. Coefficient r can vary between perfectly anti-correlated (r $=-1$) to perfectly correlated (r $=1$), with an r of 0 indicating no correlation. Upon visual inspection, we choose to define a Pearson correlation coefficient of r $\geq0.3$ as a reasonable linear correlation between extinction and emission column densities, which we deem to be more likely on the near-side of the CMZ. Clouds with  r $< 0$ are determined to be on the far-side using this method, and clouds with $0<\text{r}<0.3$ are ambiguous cases which cannot be confidently determined far-sided by this method alone. Values for each clouds are reported in Table \ref{tab: flux_calc_tab}. 

\begin{figure*}
    \centering
    \includegraphics[width=\textwidth]{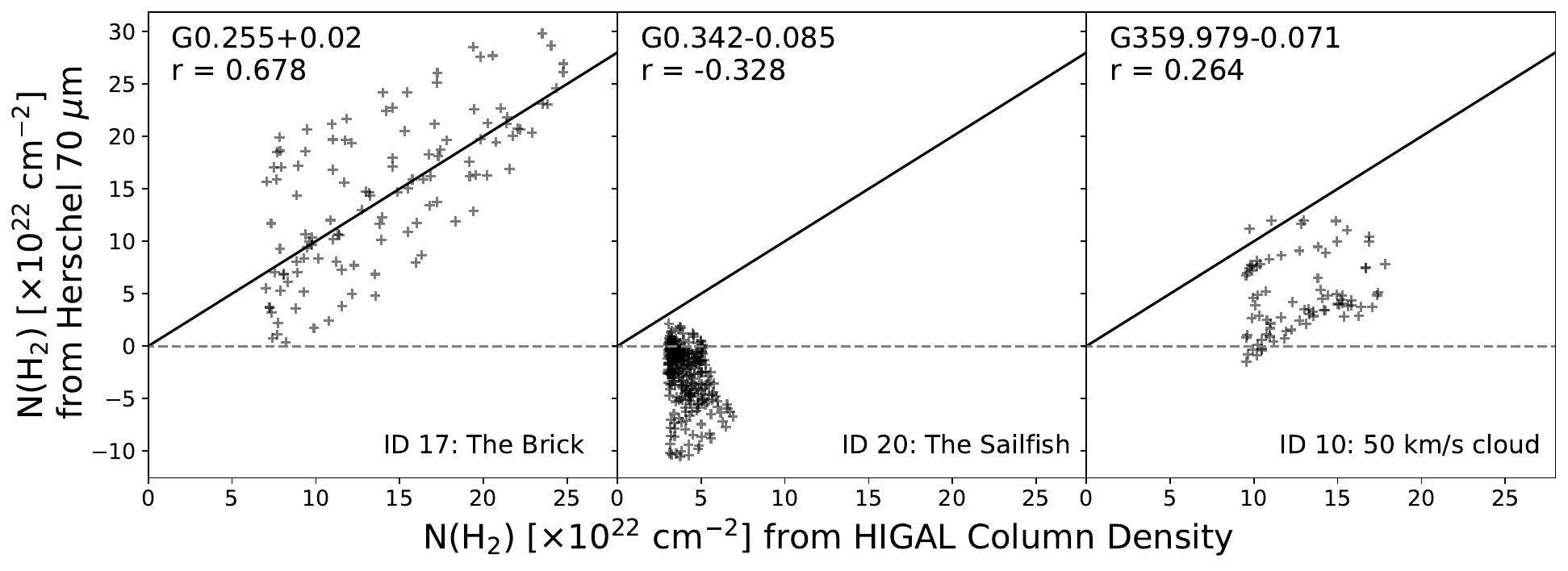}
    \caption{Herschel 70\micron calculated extinction column density compared to the emission column density from HiGAL using $f_{\rm fore}=0.5$ for three example catalog clouds: The Brick (left), The Sailfish (middle), and the 50 km/s cloud (right). The solid black line shows a 1:1 trend. A Pearson correlation coefficient of r $\geq0.3$ defines a reasonable correlation between extinction and emission column densities. We determine that reasonably correlated clouds with r $> 0.3$, such as the Brick, are likely on the near side in front of the bright galactic emission. Likewise, uncorrelated clouds with r $< 0.3$ and anti-correlated clouds with r $< 0$, are deemed to be on the far side of the CMZ. The 50 km/s cloud and the Sailfish show no correlation and negative correlation, respectively, indicative of either a lack of extinction or the presence of emission in 70\micron in the densest regions of the cloud. See \ref{fig:corr_coeff_single_comp} and \ref{fig:corr_coeff_submasks} for emission vs extinction plots for all clouds.} 
    \label{fig:corr_coeff_brick_fish_50}
\end{figure*}

A comparison of correlation coefficients for near vs far sided clouds can be seen in Figure \ref{fig:corr_coeff_brick_fish_50} for the Brick (left), the Sailfish (middle), and the 50 km/s cloud (right). We use the Sailfish as an example of clouds which appear bright in the integrated column density map, but show little extinction at 70\micron. We expect clouds like the Sailfish to show no correlation between the extinction and emission. In fact, the Sailfish shows a negative correlation between the column densities, implying the densest pixels of the cloud tend to have intensities above that of the estimated composite fore and background emission. For this method, we do not assume that clouds themselves do not emit in 70\micron, thus we interpret the negative column densities indicative of the cloud itself likely having strong emission in these areas, rather than extinction at 70\micron. Conversely, clouds with considerable extinction (i.e.\ potential near-side clouds in front of the bright galactic emission), should show some correlation. The Brick, as expected, has visibly strong extinction, and a strong correlation between extinction and emission with $r \sim 0.68$, supporting a confident placement on the near-side of the CMZ. In general, the 70\micron method estimates agree with the near/far distinctions from the flux difference and flux ratio methods.   

We further test the 70\micron method results to check that the extinction and emission in the densest parts of the cloud correlate, by imposing a threshold at the limit where the 70\micron column density becomes optically thick ($N(H_{2})_{70\mathrm{\mu m} }= 1.22\times10^{23} \mathrm{cm^{-2}}$) on both axes. Points falling within this range would correspond to opaque parts of the clouds with high extinction. There are 9 clouds that have points which fall into this range. We find a correlation coefficient for these points, $\mathrm{r}_{thick}$, which are reported in Figures \ref{fig:corr_coeff_single_comp} and \ref{fig:corr_coeff_submasks}. In all but one case where $\mathrm{r}_{thick}$ is calculated, we find that $\mathrm{r}_{thick}$ agrees with the overall near/far distinction of the total correlation coefficient $r$ (See Section \ref{sec: uncertainty_assumptions_methods}).

While largely useful in determining likely near and far-side distinctions for most clouds in the catalog, the 70\micron method still presents uncertainty that requires visual inspection (See Section \ref{sec: uncertainty_assumptions_methods}). Some clouds, such as the 50 km/s cloud, have correlation coefficients between $0<\text{r}<0.3$ which do not meet the r $=0.3$ threshold for a confident near-side distinction, but are not confidently placed on the far-side by this method alone.

Additionally, for the densest parts of CMZ clouds, the 70\micron extinction runs into similar optical depth issues as seen in typical MIR  pixel calculations still occur in the densest areas of the cloud (i.e.\ where the cloud may become optically thick), but to a much lesser extent. For example, the peak HiGAL column density in the Brick of N(H$_{2}) = 2.68\times 10^{23} \: {\rm cm^{-2}}$, corresponding to an 8\micron optical depth of $\tau_{8\mu\text{m} } = 14.66$ , and a 70\micron optical depth of $\tau_{70\mu\text{m} } = 2.18$. A detailed comparison of optical depth limitations between the 8\micron and 70\micron methods can be found in Appendix~\ref{Appendix_8vs70}.

The resulting extinction vs emission column density plots for all catalog clouds can be found in Appendix \ref{Appendix_all_cloud_corr_coeffs}. The Pearson correlation coefficient, r, for each cloud mask is reported in Table \ref{tab: flux_calc_tab}.

\subsection{Summary of Near vs Far Distinction Determinations for Individual Methods}\label{sec:results_determine_NvsF}

We determine a likely position of all CMZ clouds on either the near or far side of the Galactic Center for each of the three methods introduced in this work. Here, we remind the reader of how each method described in Section \ref{sec:new_ext_methods} is used to determine likely near/far positions for CMZ clouds.

\begin{itemize}

\item The flux difference method determines near/far side distinctions based on the average foreground, $f \sim$ 96 MJy/sr, as discussed in Section \ref{sec:flux_methods}.  Based on the average of Equations \ref{eqn: I_far} and \ref{eqn: I_near}, clouds with median flux difference values less than $\frac{CMZ}{2} - f \sim -108$ MJy/sr have higher than average foreground emission, and are more likely on the far side of the CMZ. Likewise, we determine that clouds with median flux differences greater than $\frac{CMZ}{2} - f$ are more likely on the near side of the CMZ.

\item The flux ratio method determines likely near/far positions based on each cloud's median flux ratio compared to the a flux ratio of 0.5. Thus, clouds with flux ratios less than 0.5 likely lay on the near side, while ratio greater than this number implies emission greater than that of the average foreground, placing the clouds on the far side of the CMZ.

\item The 70\micron method uses the Pearson correlation coefficient between the calculated 70\micron extinction column density and the HiGAL emission column density to determine likely near/far positions. Upon visual inspection, we choose to define a Pearson correlation coefficient of r $\geq0.3$ as a reasonable linear correlation between extinction and emission column densities. Clouds with r $\geq0.3$ are deemed to be more likely on the near-side of the CMZ, while clouds with  r $< 0$ are determined to be on the far-side using this method. Clouds with $0<\text{r}<0.3$ are ambiguous cases which cannot be determined far-side by the 70\micron method alone. 
\end{itemize}

Calculated flux difference and flux ratio values for each mask, as well as the correlation coefficient (r) are summarized in Table \ref{tab: flux_calc_tab}. We also include the measured median flux of the 8\micron image ($I_{\rm cloud}$) and smoothed model ($I_{\rm smoothed}$). In addition to the flux ratio, we calculate an ``extincted pixel fraction", $E$, defined as the fraction of pixels where the ratio $I_{\rm cloud}$/$I_{\rm smoothed}$ is less than 1. A map of completely noise would result in $E \sim 0.5$. Thus, a larger $E$ is indicative of dark clouds being ``significantly extinguished'', and more likely to be extinguished throughout, rather than having localized dark areas. Overall near/far distinctions are discussed in Section \ref{sec:results_comp_paperIII}. More detailed comparison of results is done in Section \ref{sec:discuss_NF}.

\clearpage
\setlength\LTcapwidth{\textwidth}
\startlongtable

\begin{longtable*}{ccccccccccccc}

\caption{Properties and calculated values informing the near vs far likely positions for all catalogued clouds in the CMZ. Shown for each cloud is the catalogue leaf id (clouds with multiple peak velocity features in HNCO emission were divided into separate spatial components based on the peak intensity, these separate components are denoted by letters in the leaf ids), central coordinates in degrees ($l, b$), central velocity from HNCO (v), median 8\micron flux ($I_{\rm cloud}$) in MJy/sr , median modeled emission within the cloud mask ($I_{\rm smoothed}$) in MJy/sr, flux difference in MJy/sr, flux ratio value, extinction fraction value (E), Pearson correlation coefficient between the HIGAL column density and calculated 70\micron extinction column density (r), colloquial cloud names, and reported likely positions. Clouds showing near/far positional agreement between the flux ratio and flux difference methods are noted as either near (N), likely near (LN), far (F), or likely far (LF), whereas those showing disagreement are noted as Uncertain (U).}
\label{tab: flux_calc_tab}


\\ \hline
 Leaf IDs &      l &      b &   v &  $I_{\rm cloud}$ &     $I_{\rm smoothed}$   & Flux &  Flux  &    E & r & Colloquial & Likely  \\
\newline
&&&&& & Difference & Ratio &&& Names & Positions\\ \hline

& $^{\circ}$ & $^{\circ}$ & km/s &MJy/sr & MJy/sr  & MJy/sr & &  & && N/LN/F/LF/U\\\hline

    1 & -0.525 & -0.044 & -102 &     212.67 &      317.11 &        -20.971  &     0.37  &     0.99 &    0.26 &   -          &   LN \\
    2 & -0.492 & -0.135 &  -56 &     224.21 &      276.37 &        -73.248  &     0.46  &     0.81 &    0.38 &   SgrC       &    N \\
    3 & -0.439 & -0.001 &  -90 &     197.05 &      263.15 &        -59.314  &     0.38  &     1.00 &    0.22 &   -          &    U \\
   4a & -0.405 & -0.223 &  -27 &     180.62 &      232.56 &        -79.521  &     0.37  &     0.95 &    0.43 &   -          &    N \\
   4b & -0.405 & -0.223 &  -20 &     178.70 &      234.45 &        -78.195  &     0.36  &     0.93 &    0.26 &   -          &   LN \\
    5 & -0.392 &  0.018 &  -78 &     185.01 &      230.97 &        -79.450  &     0.38  &     1.00 &    0.18 &   -          &   LN \\
   6a & -0.312 & -0.132 &  -29 &     209.38 &      285.68 &        -52.411  &     0.40  &     0.99 &    0.62 &   -          &   LN \\
   6b & -0.312 & -0.132 &  -21 &     202.07 &      259.95 &        -61.841  &     0.41  &     0.99 &    0.28 &   -          &    U \\
   7a & -0.299 &  0.032 &  -73 &     205.21 &      233.13 &        -96.863  &     0.46  &     0.78 &   -0.54 &   -          &   LN \\
   7b & -0.299 &  0.032 &  -37 &     213.06 &      255.09 &        -91.801  &     0.47  &     0.78 &   -0.62 &   -          &    U \\
   8a & -0.135 &  0.023 &  -54 &     312.12 &      325.38 &        -113.793 &     0.66  &     0.58 &   -0.05 &   -          &    F \\
   8b & -0.135 &  0.023 &  -15 &     336.89 &      305.12 &        -156.732 &     0.79  &     0.16 &      -  &   -          &    F \\
   8c & -0.135 &  0.023 &   62 &     342.50 &      310.06 &        -157.191 &     0.79  &     0.21 &   -0.64 &   -          &    F \\
    9 & -0.120 & -0.081 &   15 &     195.59 &      315.94 &        -5.056   &     0.31  &     1.00 &    0.60 &   20 km/s    &   LN \\
   10 & -0.021 & -0.071 &   48 &     314.44 &      478.22 &        38.371   &     0.46  &     0.92 &    0.26 &   50 km/s    &   LN \\
  11a &  0.014 & -0.016 &  -11 &     566.25 &      542.20 &        -148.385 &     0.87  &     0.41 &   -0.36 &   -          &    F \\
  11b &  0.014 & -0.016 &   45 &     590.78 &      522.37 &        -177.874 &     0.92  &     0.24 &   -0.35 &   -          &    F \\
  11c &  0.014 & -0.016 &   14 &     543.43 &      534.32 &        -140.236 &     0.85  &     0.36 &   -0.92 &   -          &    F \\
   12 &  0.035 &  0.032 &   86 &     495.00 &      463.88 &        -156.532 &     0.86  &     0.39 &   -0.28 &   -          &    F \\
   13 &  0.068 & -0.076 &   50 &     307.21 &      366.65 &        -65.964  &     0.58  &     0.85 &    0.14 &   Stone      &    U \\
   14 &  0.105 & -0.080 &   53 &     205.48 &      316.90 &        -13.995  &     0.34  &     0.95 &    0.50 &   Sticks     &    N \\
   15 &  0.116 &  0.003 &   52 &     477.38 &      458.67 &        -144.130 &     0.83  &     0.45 &    0.04 &   -          &    F \\
  16a &  0.143 & -0.083 &  -15 &         -  &          -  &         -       &      -    &       -  &      -  &   Straw      &   -  \\
  16b &  0.143 & -0.083 &   57 &     208.19 &      317.11 &        -16.521  &     0.35  &     0.99 &   -0.06 &   Straw      &   LN \\
  17a &  0.255 &  0.020 &   18 &     164.90 &      335.48 &        53.117   &     0.21  &     1.00 &    0.65 &   Brick      &    N \\
  17b &  0.255 &  0.020 &   37 &     170.30 &      372.12 &        74.088   &     0.20  &     1.00 &    0.72 &   Brick      &    N \\
  17c &  0.255 &  0.020 &   70 &     223.20 &      402.52 &        56.115   &     0.31  &     1.00 &    0.87 &   Brick      &   LN \\
   18 &  0.327 & -0.195 &   16 &     160.25 &      187.32 &        -98.336  &     0.34  &     0.69 &    0.10 &   -          &    U \\
   19 &  0.342 &  0.060 &   -2 &     182.75 &      250.18 &        -57.978  &     0.35  &     0.94 &    0.75 &   B          &    N \\
   20 &  0.342 & -0.085 &   90 &     299.49 &      298.69 &        -126.208 &     0.68  &     0.44 &   -0.33 &   Sailfish   &    F \\
  21a &  0.379 &  0.050 &    8 &     130.50 &      214.68 &        -41.394  &     0.16  &     1.00 &    0.54 &   C          &    N \\
  21b &  0.379 &  0.050 &   39 &     137.14 &      237.09 &        -25.267  &     0.17  &     1.00 &    0.26 &   C          &   LN \\
   22 &  0.413 &  0.048 &   19 &     114.22 &      190.32 &        -49.317  &     0.09  &     1.00 &    0.84 &   D          &    N \\
   23 &  0.488 &  0.008 &   28 &     111.19 &      197.49 &        -39.111  &     0.08  &     1.00 &    0.65 &   E/F        &    N \\
   24 &  0.645 &  0.030 &   53 &      71.23 &       99.08 &        -97.561  &     -0.25 &     1.00 &    0.61 &   -          &    N \\
   25 &  0.666 & -0.028 &   62 &      74.04 &      121.16 &        -78.284  &     -0.19 &     0.93 &   -0.33 &   SgrB2      &    U \\
  26a &  0.716 & -0.090 &   28 &      78.14 &      106.84 &        -101.027 &     -0.17 &     0.95 &    0.32 &   G0.714     &    N \\
  26b &  0.716 & -0.090 &   58 &      73.85 &       93.39 &        -105.371 &     -0.24 &     0.98 &    0.42 &   -          &    N \\
   27 &  0.816 & -0.185 &   39 &      66.86 &       75.24 &        -117.029 &     -0.39 &     0.93 &    0.10 &   -          &   -  \\
  28a &  0.888 & -0.044 &   14 &      76.41 &       87.10 &        -114.842 &     -0.23 &     0.99 &   -0.94 &   -          &   -  \\
  28b &  0.888 & -0.044 &   26 &      76.41 &       87.10 &        -114.842 &     -0.23 &     0.99 &   -0.94 &   -          &   -  \\
  28c &  0.888 & -0.044 &   84 &      72.61 &       84.80 &        -113.747 &     -0.28 &     0.98 &    0.04 &   -          &   -  \\
  29a &  1.075 & -0.049 &   74 &     144.50 &      105.39 &        -163.858 &     0.45  &     0.14 &   -0.23 &   -          &   -  \\
  29b &  1.075 & -0.049 &   85 &      84.02 &       93.97 &        -116.070 &     -0.13 &     0.80 &   -0.01 &   -          &   -  \\
  30a &  1.601 &  0.012 &   48 &      51.51 &       59.09 &        -117.923 &     -0.76 &     0.99 &    0.34 &   G1.602     &   -  \\
  30b &  1.601 &  0.012 &   58 &      52.86 &       59.99 &        -118.972 &     -0.74 &     0.98 &    0.15 &   G1.602     &   -  \\
   31 &  1.652 & -0.052 &   50 &      48.75 &       59.84 &        -114.323 &     -0.80 &     0.98 &    0.94 &   G1.651     &   -  \\

\hline
\hline

\end{longtable*}

\section{Results}\label{sec: Results}

We report our results in the following section. In Section \ref{sec:results_compare_orbital_models} we report our near/far distinctions for all CMZ clouds on four orbital models of the CMZ. In
Section \ref{sec:results_comp_paperIII} we compare the near/far results of this work with the kinematic absorption analysis from \textit{Paper III}, as well as how we determine overall near vs far distinctions summarized in Table \ref{tab: flux_calc_tab}.

\subsection{Comparison of Near vs Far Distinctions for New Extinction Techniques}\label{sec:results_compare_orbital_models}

\begin{figure*}
    \centering
    \includegraphics[width=1\textwidth]{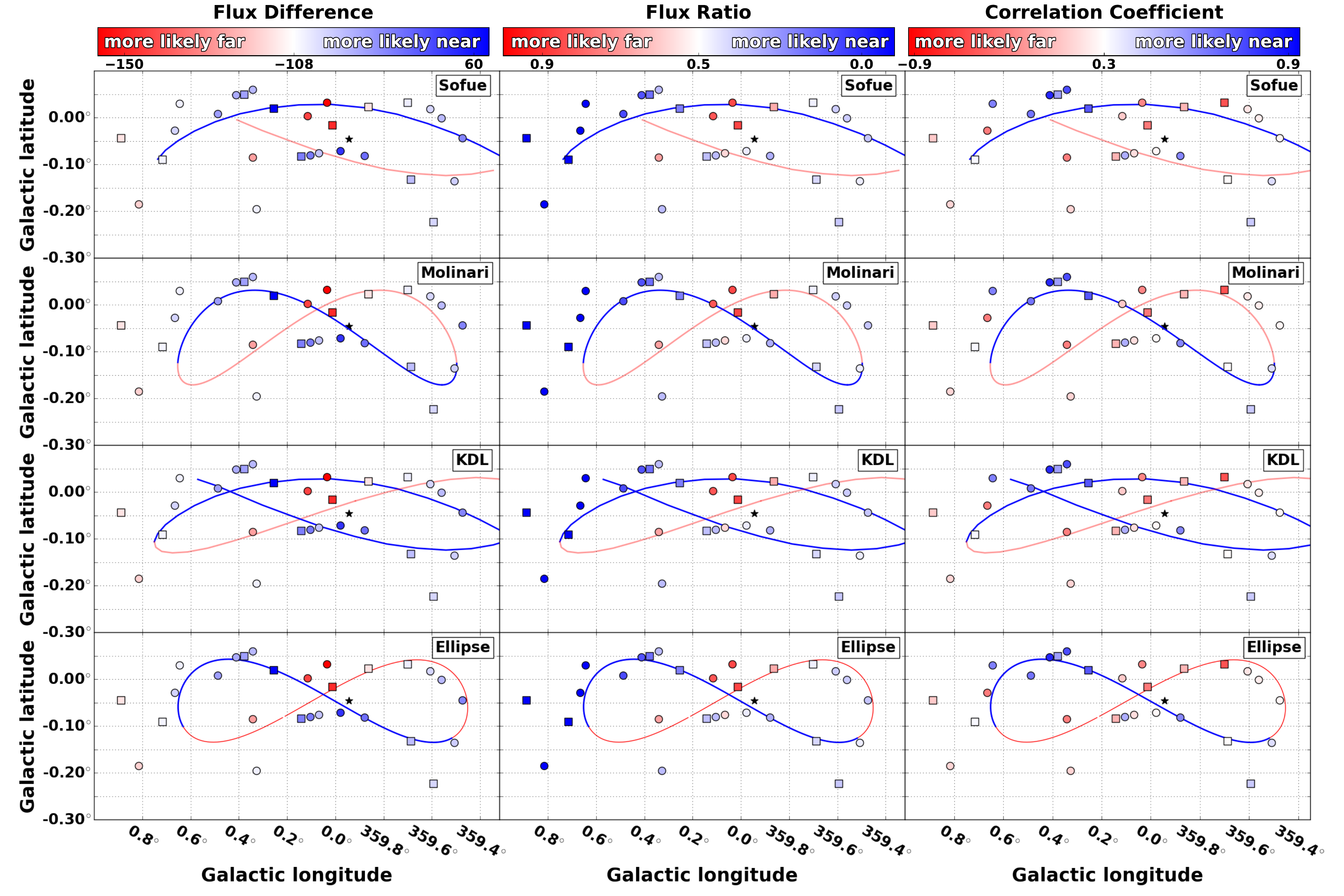}
    \caption{Comparison of flux difference (left) and flux ratio (center) methods, and the correlation coefficient based on the calculated 70\micron extinction column density compared to HiGAL N(H$_{2}$) (right) plotted over 2D projections of four orbital models in position-position space. From top to bottom: \citet{Sofue1995}, \citet{Molinari2011}, \citet{Kruijssen2015}, and a vertically oscillating elliptical orbit, similar to \citet{Molinari2011} but with conserved angular momentum in the $z$-direction and centered on ($\ell$,$b$) = (0.05\deg, -0.0462\deg). ($\mathrm{\ell}$,$b$) positions of CMZ clouds are denoted by markers. The colors correspond to the respective color bars for each method. Clouds with multiple velocity components are denoted by square markers, with the plotted color corresponding to the value of the sub-mask with the largest amount of pixels; all other clouds are denoted by circular markers. Blue colors correspond to a near-side position, while red corresponds to far side positions. The black star is the location of SgrA*. The flux difference color scale is centered on the average of Equations \ref{eqn: I_far} and \ref{eqn: I_near}. }
    \label{fig:lb_methods_combined}
\end{figure*}

We use the likely near/far distinctions for clouds determined by each of the three methods presented to distinguish between CMZ orbital models. Figure \ref{fig:lb_methods_combined} shows 2D projections of four different orbital models in position-position space, including \citet{Kruijssen2015} (KDL), \citet{Sofue1995} (Sofue), \citet{Molinari2011} (Molinari), and a vertically-oscillating elliptical orbit, similar to \citet{Molinari2011}, which assumes a constant $z$-component of the angular momentum and is offset from SgrA* at ($\ell$,$b$) = (0.05\deg, -0.0462\deg) to better fit the observed data (see \textit{Paper III} for introduction of the new ellipse model). The Sofue model presented here is based off of analysis from \citet{Henshaw2016_GasKin_250pc}, which used the open stream model to approximately trace the Sofue spiral arms in PPV space. 

\begin{figure*}
    \centering
    \includegraphics[width=1\textwidth]{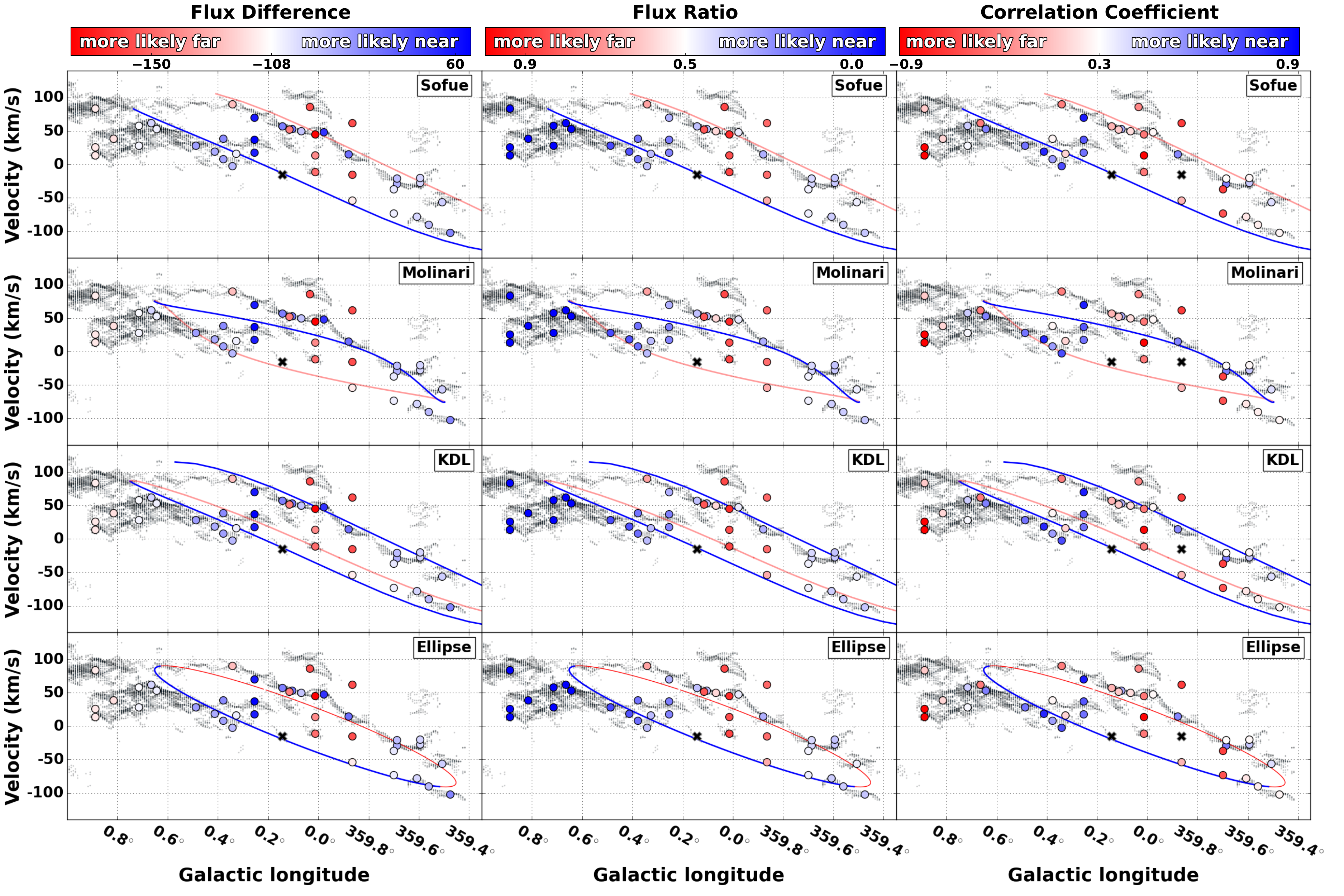}
    \caption{Comparison of flux difference (left) and flux ratio (center) methods, and the correlation coefficient based on the calculated 70\micron extinction column density compared to HiGAL N(H$_{2}$) (right) plotted over 2D projections of four orbital models in position-velocity space. From top to bottom: \citet{Sofue1995}, \citet{Molinari2011}, \citet{Kruijssen2015}, and a vertically oscillating elliptical orbit, similar to \citet{Molinari2011} but with conserved angular momentum in the $z$-direction and centered on ($\ell$,$b$) = (0.05\deg, -0.0462\deg). ($\mathrm{\ell}$,v) positions of CMZ clouds denoted as circular markers, with colors corresponding to the respective color bars for each method. Blue colors correspond to a near-side position, while red corresponds to far side positions. The flux difference color scale is centered on the average of Equations \ref{eqn: I_far} and \ref{eqn: I_near}.  Cloud component 16a, denoted by a gray cross, is an empty mask, and thus has no calculated near/far distinctions. The grey background points correspond to the spectral decomposition of MOPRA HNCO data from \citet{Henshaw2016_GasKin_250pc}.}
    \label{fig:lv_methods_combined}
\end{figure*}

For each model, the colored streams are located either in front of the GC (blue) or behind the GC (red). Likewise, the points denoting the location of catalog clouds correspond to color bars indicating the near or far position distinctions determined by the flux difference, flux ratio, and correlation coefficient; Figure \ref{fig:lb_methods_combined} left, center, and right, respectively. Blue colors indicate a likely position on the near-side of the CMZ, while red indicates a likely far-side position. Clouds with multiple velocity components are denoted by square markers, with the plotted color corresponding to the value of the sub-mask with the larger amount of pixels. Figure \ref{fig:lv_methods_combined} presents the same data and orbital models as 2D projections in position-velocity space.  

Overall, all three of the methods we introduce in this work show agreement with varying levels of confidence, particularly for clouds on the 100~pc ring, between longitude ranges -0.5\deg $< \ell <$ 0.7\deg. All four methods tend to place more clouds on the near side of the CMZ, which agree best with the ($\ell$, $b$) projections for the Molinari, KDL, and Ellipse orbital models. Similarly, the results for each method in ($\ell$, v) space seem to agree best with the KDL or ellipse model. Despite some variation between methods for the near/far results in position-position and position-velocity space, all three new methods presented produce results which appear to fit each orbital model with the similar levels of confidence. We present both qualitative and quantitative fits to each orbital model in depth in Sections \ref{sec:discuss_3Dmodels_qualitative}  and \ref{sec:discuss_3Dmodels_quantitative}, respectively.

\subsection{Comparison and Synthesis with Absorption Data}\label{sec:results_comp_paperIII}

In addition to comparing the results of our three dust extinction methods to each other, we compare our results to independently determined near/far distinctions from the analysis in \textit{Paper III}, which took a ratio of 4.8~GHz radio continuum emission over measured line absorption from H$_{2}$CO (1$_{1,0}$ - 1$_{1,1}$) data to estimate near/far positions of the clouds in the catalog. The line absorption analysis uses a radio continuum-absorption ratio of 1.0 as a threshold, where a ratio less than 1.0 indicates a near-side distinction, and a ratio greater than 1.0 denotes far-side.

Figure \ref{fig:compare_methods_3panel} shows comparisons of likely near vs far positions between the flux difference and the other near/far distinguishing methods from this work and \textit{Paper III}: the flux ratio (top panel), 70\micron correlation coefficient (middle panel), and molecular line absorption from \textit{Paper III} (bottom panel). The blue shaded areas in each plot denote ranges where the methods agree on a near-side distinction, whereas the red shaded areas are regions which show far-side agreement for the compared methods in each plot.  The flux difference and flux ratio show the most agreement with the least spread, and minor discrepancies between nine leaf IDs which lie on the edge or off of the CMZ ring (i.e. 19\% disagreement). The correlation coefficient and absorption analysis show slightly larger spreads when compared to the flux difference, with 31\% and 32\% disagreement, respectively. The location of disagreements between these methods and the flux difference are not limited to a particular region of the CMZ. Although, most of the near/far discrepancies are quite minor, particularly between the flux difference and correlation coefficient results. Overall, all four methods show good agreement between each other, with varying levels of confidence.

In order to summarize the overall likely positions of each CMZ catalog cloud from this paper series, we do a visual inspection of the four methods summarized in Figure \ref{fig:compare_methods_3panel} to determine a near or far likely position for each cloud. Our overall near/far distinctions are reported in the ``Likely Positions" column of Table \ref{tab: flux_calc_tab}. IDs which are placed confidently on the near side in all four methods are deemed ``Near" (N). Similarly, clouds which are confidently far side in all methods are deemed ``Far" (F). Some clouds show confident positions in two or three methods, but are either missing absorption data or have ambiguous, yet conflicting, near/far distinctions between methods. In these cases, we make a visual determination for whether the cloud is ``Likely Near" (LN), ``Likely Far" (LF), or has an ``Uncertain" (U) position based on the methods in our analysis. We note that some of the near/far classifications we make here would not be applicable for clouds which are not on the standard x2 orbit of the CMZ (See Section~\ref{sec: uncertainty_assumptions_morph}).

Relevant data products and code used for the analysis from this paper series have been made publicly available via github: \url{https://centralmolecularzone.github.io/3D_CMZ/}. The maps and data products can be found on Harvard DataVerse (DOI: 10.7910/DVN/FBV7T5): \url{https://dataverse.harvard.edu/dataverse/3D_CMZ}

\section{Discussion}\label{sec: discussion}

We discuss the results in the following sections. Section \ref{sec:discuss_NF} covers the most notable areas of agreement between methods, as well as major disagreements and minor discrepancies. In Section \ref{sec:discuss_3Dmodels_qualitative} we provide a qualitative discussion of the combined results from all three of the methods introduced in this paper, as well as the absorption analysis from \textit{Paper III} and their overall fit of the summarized likely near/far positions of CMZ clouds on the four orbital models previously discussed. In Section \ref{sec:discuss_3Dmodels_quantitative}, we present a quantitative percent agreement of the combined data in $lbv$ space to each of the theoretical models. Finally, in Section \ref{sec: uncertainty_assumptions_methods} we note the uncertainties and limitations of the methods presented above.

\subsection{Comparison of methods for distinguishing near and far locations}\label{sec:discuss_NF}

\begin{figure}
    \centering
    \includegraphics[width=0.42\textwidth]{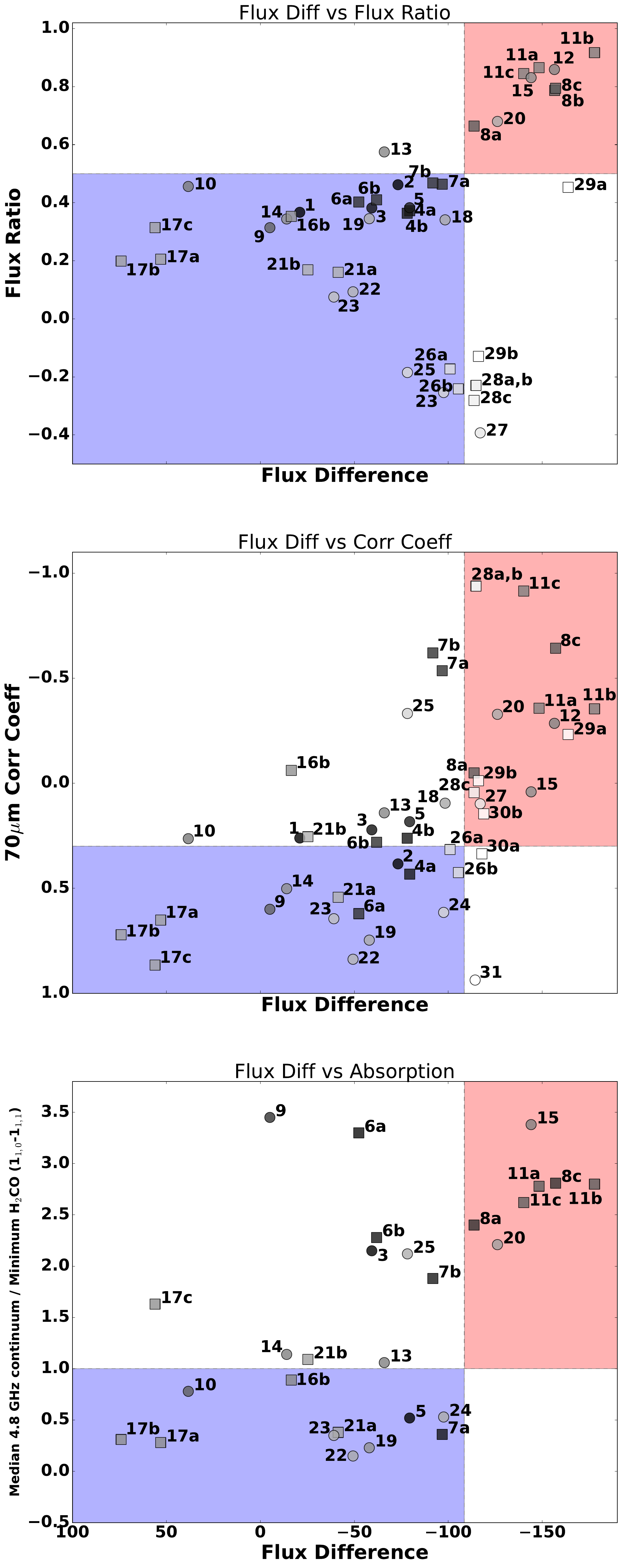}
    \caption{Comparison of likely near/far positions between the flux difference and three other methods, including the flux ratio (top), 70\micron correlation coefficient (middle), and absorption measures from \textit{Paper III} (bottom).\ Clouds are noted by circular markers with annotated IDs. Marker colors correspond to the positions of clouds at more negative longitudes (darker) or more positive longitudes (whiter). Square markers denote clouds with multiple measured velocity components. Clouds within shaded colored regions show agreement of likely near (blue) or far (red) positions between methods.}
    \label{fig:compare_methods_3panel}
\end{figure}

In this Section we discuss the most notable areas of agreement between methods, major disagreements, and minor discrepancies.

The three methods discussed in this paper show overall agreement for CMZ ring clouds which lie within longitude ranges -0.5\deg $< \ell <$ 0.7\deg (between IDs 1 to 26). The Clouds on the CMZ ring between SgrB2 and the Brick, often referred to as the dust ridge, have previously been assumed to lie in the foreground of the CMZ, with the strongest line-of-sight positional constraints from proper motions placing SgrB2 in the foreground at about 0.13 kpc in front of the Galactic Center \citep{Reid2009}. The flux difference and flux ratio methods confidently place all of the clouds lying on the contiguous dust ridge stream in the foreground. Additionally, all of the methods compared in Figure \ref{fig:compare_methods_3panel} show strong agreement on far-side determinations for IDs 8,11,12, and 15, which lie close in longitude to SgrA*. However, the bright emission around SgrA* makes confidently determining near or far positions of these clouds difficult, as much of the bright emission could be due to the background, rather than the clouds themselves.

Major disagreements between methods occur at positive longitudes past the dust ridge, where clouds lie either on the edge or off the CMZ ring for each of the geometric models. While the flux ratio places all of the positive longitude clouds on the near side of the CMZ, the flux difference and 70\micron correlation coefficient methods show more ambiguous far-side distinctions. For example, leaves 30a, 30b, and 31 lie far off of the CMZ ring and are determined confidently near-side by the flux ratio, but are deemed far side by the flux difference (see Figure \ref{fig:flux_methods_full_maps}), and have either uncertain or confident nearside distinctions based on the 70\micron method. Determining the positions of all clouds that lie off the CMZ ring (at $\ell \gtrsim$ 0.7\deg) is complicated due to the inhomogeneous nature of the ISM, and the ambiguous source of background emission becoming a more important factor off of the orbital streams. Additionally, the lack of molecular line absorption data for these clouds (see \textit{Paper III}) makes an overall near vs far distinction uncertain with the available data and methods for this paper series. 

Minor discrepancies between the three extinction methods occur near SgrA* at longitudes $ -0.2$\deg $ < \ell < 0.2$\deg. Our methods seem to agree overall, but with varying degrees of confidence. In particular, the controversially discussed  50 km/s cloud (ID 10) \citep{Ferrire2012} appears to be on the near side based on the flux difference and flux ratio, but is placed likely on the far-side based on the 70\micron method and absorption data. However, \textit{Paper III} concluded the 50 km/s cloud lies on the near side, as its absorption spectrum is too deep to place it on the far side. Taking a closer look at the 50 km/s cloud's emission vs extinction (Figure \ref{fig:corr_coeff_single_comp}), the higher HiGAL column density values for the cloud appear to saturate in the extinction. We found that most of the lower extinction column density points for the cloud occur for areas closer to SgrA*. It is possible that the strong emission from SgrA* heavily skews the extinction calculation for clouds in this region, and thus impacts the 70\micron method near/far distinctions for clouds near SgrA*. It may also be possible that these clouds actually lie closer to the circumnuclear disk (CND) orbiting at a radius of $\sim$10~pc from SgrA*,  and not necessarily on the near or far side of the CMZ ring itself. This idea has been suggested and supported by simulations from \citet{Tress2020}, which proposed the 20 and 50 km/s clouds may lie closer to SgrA*, based on their morphological and kinematic connections to gas structures in the CND \citep[e.g.][]{Tsuboi2018,Ballone2019}, and their high column densities despite their lack of extinction features in the 8\micron, as seen in this study. However, the current methods and observations lack sufficient information and resolution to either support or refute this possibility. Additionally, time-averaged simulations may deviate significantly from any particular snapshot of the present-day CMZ. The complexity of the gas structures and emission in the central region near SgrA* suggests that current models are too oversimplified in the context of the current CMZ.

Overall, the three near/far distinction methods presented in this paper (flux difference, flux ratio, and 70\micron correlation) show good agreement. The molecular line absorption method from \textit{Paper III} is completely independent of the methods presented here, yet still shows good agreement in near/far distinctions. This paper's 8\micron and 70\micron methods are independent of each other, but they employ the same general assumptions of constant foreground and background distributions compared to the highly variable distribution in the CMZ region. The flux ratio and flux difference methods also assume CMZ clouds completely absorb the 8\micron background emission, which may be invalid in areas impacted by the emission from SgrA*. 

\subsection{Qualitative Comparison to 3-D Models of the CMZ}\label{sec:discuss_3Dmodels_qualitative}

Our determinations of likely near/far positions for individual molecular clouds provides new evidence necessary to help distinguish between current theoretical orbital models of the CMZ. Figure \ref{fig:LB_megaplot} shows the ($\mathrm{\ell}$,$b$) projection of CMZ clouds over four geometric orbital models, colored according to their summarized likely near vs far positions from the four methods presented in this work and \textit{Paper III}. The colors correspond to Near (blue), Likely Near (cyan), Likely Far (rose), and Far (red) positions. Clouds that show substantial disagreement between methods are marked Uncertain (gray). We do not show clouds past $\ell > \sim 0.7$\deg that lie off of the 100~pc stream. Figure \ref{fig:LV_megaplot} shows the same summarized near vs far locations in ($\mathrm{\ell}$,v) space. The horizontal extent of each ($\mathrm{\ell}$,v) point corresponds to the projected radius recorded in the dendrogram catalog table, while the vertical extent corresponds to the velocity dispersion.

The near/far alignment of clouds onto the ($\ell$, $b$) and ($\ell$, v) projections of orbits can help compare the potential fit of each model. We note that while some regions show good agreement with our summarized near/far distinctions for all models, no single model fits the results perfectly. A perfect agreement is not expected, given the complex gas flows in the CMZ. 

The Sofue spiral arms show the clearest disagreement in position-position space, while the Molinari, Ellipse, and KDL streams models provide good fits to the summarized results in ($\mathrm{\ell}$,$b$) space. However, the Molinari closed elliptical orbit has a notable disagreement in ($\mathrm{\ell}$,v) space along the dust ridge clouds between SgrB2 and the Brick, which are largely considered near-side clouds in front of the bright Galactic emission. All other models presented in this paper series show a better ($\mathrm{\ell}$,v) fit to the confidently near-sided dust ridge, particularly in position-velocity space.

Clouds 3 and 5 lie along a stream of material between $ -0.65$\deg $ > \ell > -0.0$\deg and $ -0.05$\deg $ > b > -0.1$\deg. This region is often colloquially referred to as the ``wiggles" region, as this area's formation is likely driven by gravitational instabilities leading to periodic density and velocity variations along the stream \citep{Henshaw2020,Henshaw2016_Seeding}. \citet{Henshaw2016_GasKin_250pc} argue these clouds are likely on the near side, which would favor the KDL geometry. However, \citet{Kruijssen2015} state that near/far side placement of these clouds does not change their best-fitting orbital solution. While the KDL open stream shows consistent agreement between ($\ell$, $b$) and ($\ell$, v) projections, it does not provide an explanation for how inflowing material connects to the CMZ orbit, and thus lacks a connection to large-scale dynamics.

Similar to the KDL streams, the toy Ellipse model shows good agreement to the near/far distinctions in both ($\ell$, $b$) and ($\ell$, v) projections. In particular, the Ellipse model agrees with the confident near-side placement of the dust ridge clouds in position-velocity space, which the Molinari ellipse fails to reconcile. We note that the agreement of near/far positions from all four distinction methods with the Ellipse model in this paper differs slightly from the results in \textit{Paper III}. However, both analyses support the potential for an elliptical x2 orbit to provide a good fit to the data. The two elliptical models follow the interpretation of of \citet{Binney1991}, which characterized the dense gas in the CMZ as following an elliptical orbit perpendicular to the major axis of the Galactic bar that transports inflowing material towards the CMZ. However, both elliptical models are empirical fits to the observed data and still require physical modeling to determine the most probable solution for a closed elliptical orbit.

\subsection{Quantitative Comparison to 3-D Models of the CMZ}\label{sec:discuss_3Dmodels_quantitative}
In order to provide a quantitative measure to determine a "best fitting" orbital model with the current combined data, we present multiple methods to assess how well each model matches the data in both a 3-D positional ($\ell bv$) space as well as in their near vs far classifications. In this paper, we present a simplistic minimum-distance approach to determine if a cloud lies close to a given model's orbital streams, and then assess the agreement with stream's near/far classification. 

To start, we resample the orbital models such that they lie on a common $\ell bv$ grid with the same amount of sampling points, each with a near or far classification. The ($\ell$, $b$, v) positions of each model point and catalog cloud are then normalized on a scale of 0 to 1, using a min-max normalization procedure:

\begin{equation}
    z = (x - \mathrm{min})/(\mathrm{max} - \mathrm{min})
\end{equation}\\

Where the normalized value $z$ of a given variable $x$ is determined by the minimum and maximum values of the dataset. We use approximate extents of the data in each coordinate axis. Normalizing each $lbv$ coordinate on a scale of 0 to 1 ensures we are able to directly compare minimum distances between cloud positions and model points in 3-D position-position-velocity space. 

Using the normalized model and data positions, we separately determine the percent agreement for a given model to each cloud in the catalog. We calculate the Euclidean distance between a cloud's normalized $lbv$ position to a the model's sampled points, and identify the minimum distance between them. If the minimum distance falls within a threshold of 0.25 (in normalized $lbv$ space), it is considered a positional ``match". Lastly, we assess the near/far agreement of the cloud to the stream on which its minimum distance placed it. If the near/far distinction for the stream ($\mathrm{NF}_{\mathrm{stream}}$) agrees with that of the cloud ($\mathrm{NF}_{\mathrm{cloud}}$), it is considered a match and added to the percent agreement for that model.

In other words, the Euclidean distance-based percent agreement is the percentage of clouds which satisfy both of the following conditions:

\begin{equation}
  \mathrm{min~distance} = \sqrt{\Delta z_l^2 + \Delta z_b^2 + \Delta z_v^2} < 0.25
\end{equation}

\begin{equation}
    \mathrm{NF}_{\mathrm{stream}} = \mathrm{NF}_{\mathrm{cloud}}
\end{equation}\\

Where $\Delta z_l^2$, $\Delta z_b^2$, and $\Delta z_v^2$ are the differences between the normalized $lbv$ model points and clouds coordinates. The threshold of 0.25 was arbitrarily chosen to reasonably agree with the a by-eye determination. The $lbv$ fitting is sensitive to the threshold value, which can significantly impact the spatial-velocity fit for all models. However, the overall agreement is, as expected, most impacted by the N/F agreements between the clouds and the model streams.

The results of the minimum distance agreement approach are summarized in the second column of Table \ref{tab:model_agreements}, and visually represented in separate ($\ell$, $b$) and ($\ell$, v) space in Figure \ref{fig:percent_agree}. We find that the KDL model provides the best near/far agreement in $lbv$ space (65\% agreement), although it is closely matched by the toy Ellipse model (61\%), followed by the Sofue spiral arms (55\%) and the Molinari ellipse (23\%). 

The main disparity between the KDL and Ellipse models appears in the upper right quadrant in ($\ell$, $b$) space near the ``wiggles" region, as discussed in the previous section. The KDL model agrees with the combined near-sided placement of these clouds. However, the region lies on the edge of a far-sided stream in the Ellipse model. Additionally, the Ellipse fails to place the 20 and 50 km/s clouds on the near stream based on their positions in velocity space. The Sofue spiral arms, while in agreement with the near-side placement of the ``wiggles", also fails to match with the points near SgrA* in ($\ell$, v) space. 

The Molinari model performs the worst out of the four by a considerable difference. As seen in the qualitative analysis, the Molinari ellipse does not agree with near/far positions for multiple regions in ($\ell$, v) space, including many of the confidently near-sided dust ridge clouds, the ``wiggles" region, or the central Sgr A* region. Thus, despite the Molinari model's good fit in the ($\ell$, $b$) projection, it is not able to recover the clouds' positions in velocity space, and is heavily penalized for that in the minimum distance fitting.

In addition to the Euclidean minimum distance-based approach described here, we also report results from an alternative agreement evaluation method utilizing a k-nearest neighbors (kNN) approach presented in detail in \textit{Paper III}. There are many subtle differences in the normalization and distance measures which differentiate this methods from a straightforward minimum distance approach. In brief, the kNN agreement method involves identical preprocessing of the models as described in this work. The data points for the clouds are similarly normalized on a scale from 0 (near) to 1 (far), using the \texttt{RobustScaler} from \texttt{scikit-learn}. Unlike the scaling used for the minimum distance method, the kNN method's normalization centers the data by subtracting the median and scales it according to the interquartile range. The minimum $lbv$ distance between the normalized points and the models are found using the Mahalanobis distance metric, which is better able to identify outliers than the Euclidean distance. Lastly, \texttt{scitkit-learn}'s \texttt{NearestNeighbors} functionality is used to assess the near/far agreement of each point with its matching position on the model streams, using a number of neighbors, k, determined by the square root of the number of model points, N (i.e. $ k = \sqrt{N}$). The near/far assignment of the model is then based on a weighted vote of the near/far positions of the neighboring model points in $lbv$ space. The results from the kNN approach are summarized in the third column of Table \ref{tab:model_agreements}. The kNN approach finds a similar ranking in model preference: KDL (69\%), Sofue (61\%), Ellipse (60\%), and Molinari (55\%). We direct the reader to \textit{Paper III} for a detailed description of the kNN agreement method. 

Both quantitative agreement methods place the KDL model first and Molinari last (albeit with varying degrees of agreement). The top three ranked models (KDL, Ellipse, and Sofue) are all fairly close in agreement, within 10\% of each other using both methods. However, the percent agreement for the Molinari model is more heavily penalized by the minimum distance approach compared to the kNN method. This is likely due to the kNN method lacking a strict minimum threshold to declare if a point is close enough to the streams to be considered a ``match" before near/far assessment. The lack of an $lbv$ threshold means the kNN method does not necessarily measure if a given model matches the data well, but rather focuses on ranking which models perform better than others. The minimum distance approach's $lbv$ threshold directly accounts for whether the models themselves accurately describe the spatial-velocity distribution of the clouds. Additionally, both the minimum distance and kNN agreement methods weight the $lbv$ and near/far assessment equally, and do not fold in prior information or confidence concerning the positions of certain clouds (e.g. the confident near positions of the dust ridge clouds). We present the methods here as multiple ways to quantitatively determine model rankings. The creation of a more thorough and statistical fitting analysis is planned for a future paper.

\begin{table}
\centering
\caption{Quantitative comparison of orbital models using a minimum-distance approach, resulting in a percent agreement (described in Section \ref{sec:discuss_3Dmodels_quantitative}), or a K-nearest neighbors approach (detailed in \textit{Paper III}), resulting in near/far accuracy score. Both methods assess each cloud's minimum distance to the orbital model streams, and then evaluate whether its near/far distinction agrees with the model stream's near/far position.}
\label{tab:model_agreements}
\begin{tabular}{lccc}
\hline
Model & \% agreement & kNN score \\
 & (min distance method) & (method from \textit{Paper III})  \\
\hline
KDL & 65\% &  69\% \\
Ellipse & 61\% &  61\% \\
Sofue & 55\% &60\% \\
Molinari & 23\% & 55\% \\
\hline
\end{tabular}
\end{table}

Our quantitative analysis concludes that none of the present 3-D CMZ orbital models definitively fit the observed emission in both position-position and position-velocity space. There are also five clouds with uncertain near/far positions based on our results: IDs 3, 5, 6, 13 (Stone), 18, and 25 (SgrB2) for which more work is needed in order to confidently determine their near/far positions. The uncertainty in distinguishing the best fitting models based on individual regions makes it important to have confident near/far placements for as many structures on the CMZ ring as possible. However, it is likely these current models are too simplistic to accurately describe the orbital motions of gas in the CMZ. Thus, a more complex, dynamically driven model may be necessary.

\subsection{Uncertainties and Assumptions}
In this section, we address the assumptions and uncertainties pertaining to each method presented above (\ref{sec: uncertainty_assumptions_methods}), as well as regarding the currently assumed CMZ morphology (\ref{sec: uncertainty_assumptions_morph}).

\subsubsection{Limitations of the New Dust Extinction Methods}\label{sec: uncertainty_assumptions_methods}

Both the flux difference and flux ratio methods assume a constant background and constant foreground, implying all observed variability comes from the CMZ area. This is a broad simplification we make due to the observed, extremely bright emission at 8\micron spanning from the central area around SgrA* to the outer edge of the CMZ ring. The flux difference and flux ratio methods show overall good agreement with each other, with the most prominent disagreements occurring at the positive longitude edges of the CMZ ring. The flux ratio places clouds at $\ell > 0.5^\circ$ on the near-side, while the flux difference shows more ambiguous near or far positions, particularly past SgrB2. This disagreement could in part be due to the more diffuse background emission outside of the main CMZ ring ($\ell > 0.7$\deg) compared to the bright central region nearer to SgrA*; the clouds in this region may be extinguishing light from behind, but the diffuse emission makes it difficult to compare the relative ``darkness" of the clouds to the background as a means of measuring absorption. Additionally, outside of the CMZ ring, the basic schematic presented in Figure \ref{fig:flux_diff_schematic} does not hold true, and it is much less certain where the bright illumination behind the clouds actually occurs. 

The 8\micron methods also show agreement with the 70\micron correlation coefficient results. The modified 70\micron correlation for the dust ridge clouds shows the calculated extinction and emission column densities map onto the 1:1 trend line quite well, as expected for the typical 8\micron version of this method for IRDCs in the Galactic disk (see Figures \ref{fig:corr_coeff_single_comp} and \ref{fig:corr_coeff_submasks}). However, the correlation coefficient and absorption analysis place SgrB2 in the background. Interestingly, the optically thick limited correlation coefficient for SgrB2, $\mathrm{r}_{thick}$, places the areas of the cloud that best correlate the densest and darkest material as "uncertain" rather than in the background. The near/far distinctions for SgrB2 could be strongly impacted by the region's high rate of embedded high-mass star formation (particularly for the 8\micron and 70\micron methods), or even by artifacts in the ATCA line data near this region (as noted in \textit{Paper III}, which concludes SgrB2 must be on the near side). For example, the 70\micron extinction column density correlation, which relies on the correlation of pixel-by-pixel comparisons of the cloud's emission to the smoothed background model, is strongly skewed by these bright embedded sources, and possibly more extended diffuse emission from the clouds themselves, that result in very low calculated extinction values compared to the column density, regardless of the value of f$_{\rm fore}$. The flux difference and flux ratio methods instead only take into account the intensity of the observed emission compared to the modeled background. We note that the correlation coefficient method could be improved by incorporating a distribution of the 70\micron emission in the disk (i.e. calculating \ffore\xspace for each source separately). Such a model could potentially make it possible to determine line of sight distances to clouds to some degree. However, there are multiple mechanisms that majorly contribute to the 70\micron emission throughout the CMZ. Confidently modeling the 70\micron distribution in the disk is difficult without the use of carefully constructed radiative transfer models, which is outside the scope of this paper.

While we mostly use our newly presented dust extinction methods as a measure of near/far likelihood, these methods could potentially be used to compare the ``relative depth'' of clouds through the CMZ, though with some caution. This is probably best used to compare the relative depth of clouds that are near each other in projection (e.g. comparison of the Three Pigs, or perhaps the dust ridge clouds). Uncertainties in the methods increase closer to SgrA* in projection due to the bright emission from the GC, as well as towards the edge of the CMZ due to the variable ISM. So the methods are best used for a near/far positional confidence, though the depth interpretation could reasonably be used on local scales.

\subsubsection{What do we define as the CMZ?}\label{sec: uncertainty_assumptions_morph}

It is important to note there exists uncertainty regarding what is generally defined as the ``CMZ ring", as well as what features are assumed to lie on the ring. The models presented here, and in previous studies in the literature, often aim to fit the orbit of dense gas based on the observed twisted infinity shape of the ring. However, assuming that all of the structures identified should lie on the 100~pc ring is sometimes a point of controversy. The 20 and 50 km/s clouds, for example, are often deemed as potential candidates for in-falling material onto the CND based on their line-of-sight positions and velocities \citep{Lee2008,Karlsson2003}, or may reside much closer to SgrA* than the rest of the gas on the ring \citep{Yan2017}. While the clouds do pass through the velocity range of the CND (-100 to +100 km/s), however, it does not necessarily mean the clouds are directly connected to that material. Simulations from \citet{Tress2020} offer an alternative interpretation, that the clouds could be filling the space between the 100~pc ring and the CND. In general, the ambiguous connection of the 20 and 50 km/s clouds to the 100~pc ring or the CND often obscures typical model fittings. If the clouds do indeed fill the space between the ring and the CND, they should not necessarily be used to assess the orbital models or the outer orbits that most of the dense gas follows. 

It is also worth considering the extreme case where all of the clouds exist in a fluid range of distances throughout the CMZ, not necessarily on the orbital streams. In this case, the use of the methods as distance proxies (i.e. relative depths of clouds in the CMZ) become more valuable than the near/far distinctions alone. This paper focuses on simple geometries that mostly attempt to describe gas on a single ring-like orbit, where the binary near/far positions are helpful to distinguish the models. However, the underlying measurements and approaches are valuable for measuring more complex geometries, albeit with important systematics.

The extent of the CMZ has also been widely questioned. As previously noted, the models presented here aim to fit orbits primarily to structures within the central 100~pc (though \citet{Sofue1995} and \citet{Sofue2022} also fit the spiral arms model to structures that lie off of the typical extent of the ring). MHD simulations utilizing Milky Way-like Galactic potentials often result in CMZs with radii of 300-400~pc \citep{Sormani2020,Tress2020,Hatchfield2021,Tress2024}. Additionally, observations, including those used for this series, have shown dense gas that is often associated with the CMZ lying at longitudinal extents far from the typical orbital streams \citep{Ginsburg2016,Purcell2012,Rigby2016_CHIMPS}. In fact, Figure 4 of \textit{Paper I} in the this series presents the extent of the CMZ to lie between $-1.3$\deg$<|\ell|<1.8$\deg, based on the dust temperatures and dense gas profile for the region as observed with Herschel. Thus, it may be falsely limiting to assume the models must only fit the central 100~pc, or that all objects within that region should definitely lie on the ring (e.g. 20 and 50 km/s clouds). Detailed testing and consideration of different CMZ sizes, and weighting of certain structures' fits when developing models are planned for a future paper in this series.

\section{Summary and Conclusions}\label{sec:summary_conclusions}

We investigate the application of three independent dust extinction methods to CMZ catalog clouds in order to determine their likely position either on the near or far side of the Galactic Center: 1) the 8\micron flux difference, 2) 8\micron flux ratio (Section \ref{sec:flux_methods}); and 3) the correlation between 70\micron extinction and emission column densities (Section \ref{sec:70um_ext_method}). The relevant code for this project are publicly available via: \url{https://centralmolecularzone.github.io/3D_CMZ/}. The maps and data products can be found on Harvard DataVerse (DOI: 10.7910/DVN/FBV7T5): \url{https://dataverse.harvard.edu/dataverse/3D_CMZ}

We summarize our key new methods and main findings below.

Our main result is a systematic measurement of the relative line-of-sight positions of clouds in the CMZ. These data are useful for testing models of gas orbits in the CMZ, and provide a basis for testing future modifications or additions to current CMZ orbital models. 

\begin{figure*}
    \centering
    \includegraphics[height=\linewidth]{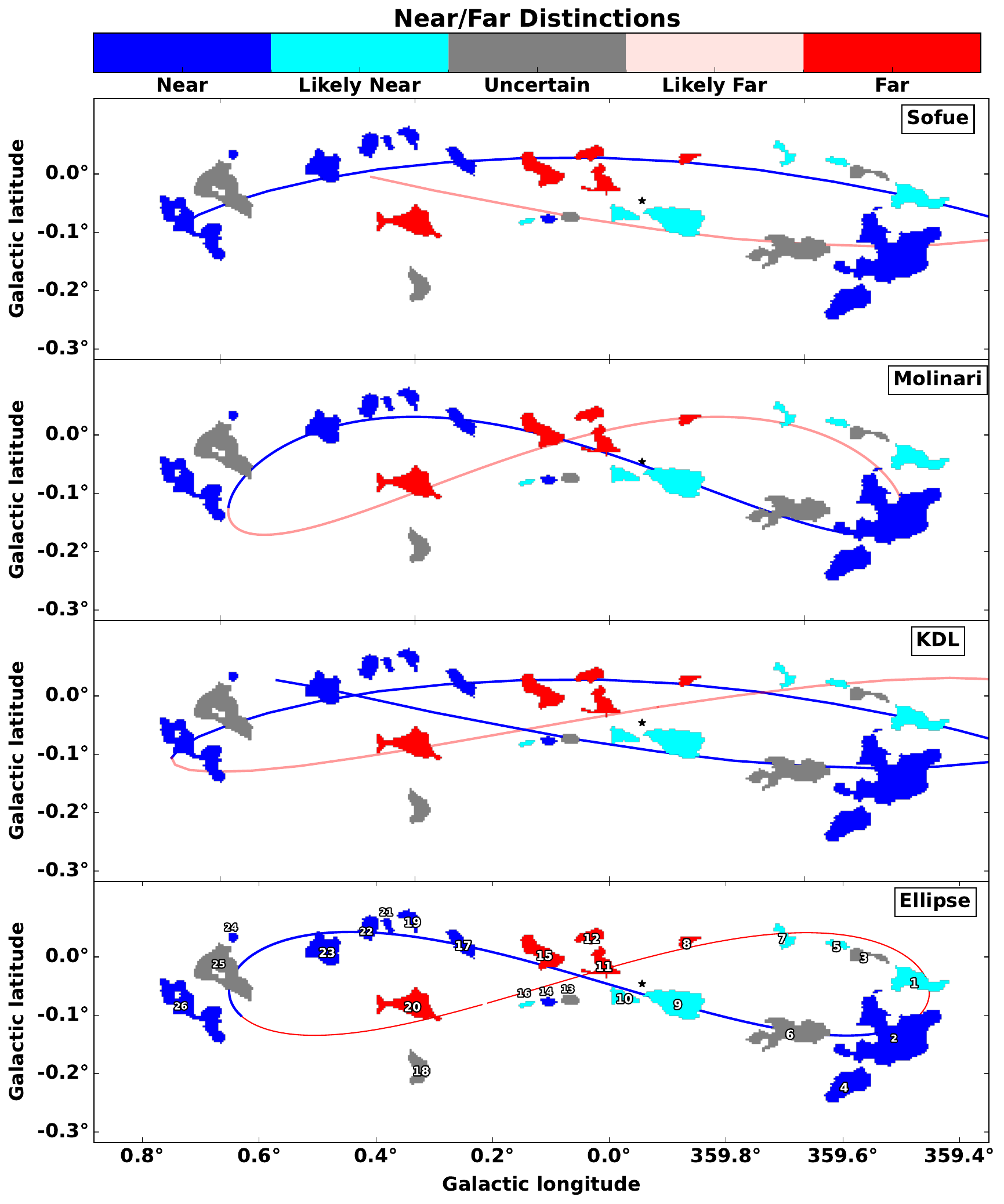}
    \caption{($\mathrm{\ell}$,$b$) projection of summarized near vs far likely positions of clouds in the CMZ ring based on comparison of the flux difference, flux ratio, 70\micron correlation coefficient, and absorption analysis. Catalog masks are plotted over the 2D projections of four orbital models in position-position space. From top to bottom: \citet{Sofue1995}, \citet{Molinari2011}, \citet{Kruijssen2015}, and a vertically oscillating elliptical orbit, similar to \citet{Molinari2011}. The colors correspond to Near (blue), Likely Near (cyan), Likely Far (rose), and Far (red) positions of clouds based on visual summarizing of the four methods from Figure \ref{fig:compare_methods_3panel}. Clouds that show disagreement between two or more methods are marked Uncertain (gray). The black star notes the location of SgrA*. We do not show clouds past $\ell > \sim 0.7$\deg that lie off the main CMZ ring. The elliptical or KDL streams models show the best fits to the current summarized results, though it is not possible to confidently distinguish between them with the given data and methods.}
    \label{fig:LB_megaplot} 
\end{figure*}

\begin{figure*}
    \centering
    \includegraphics[height=\linewidth]{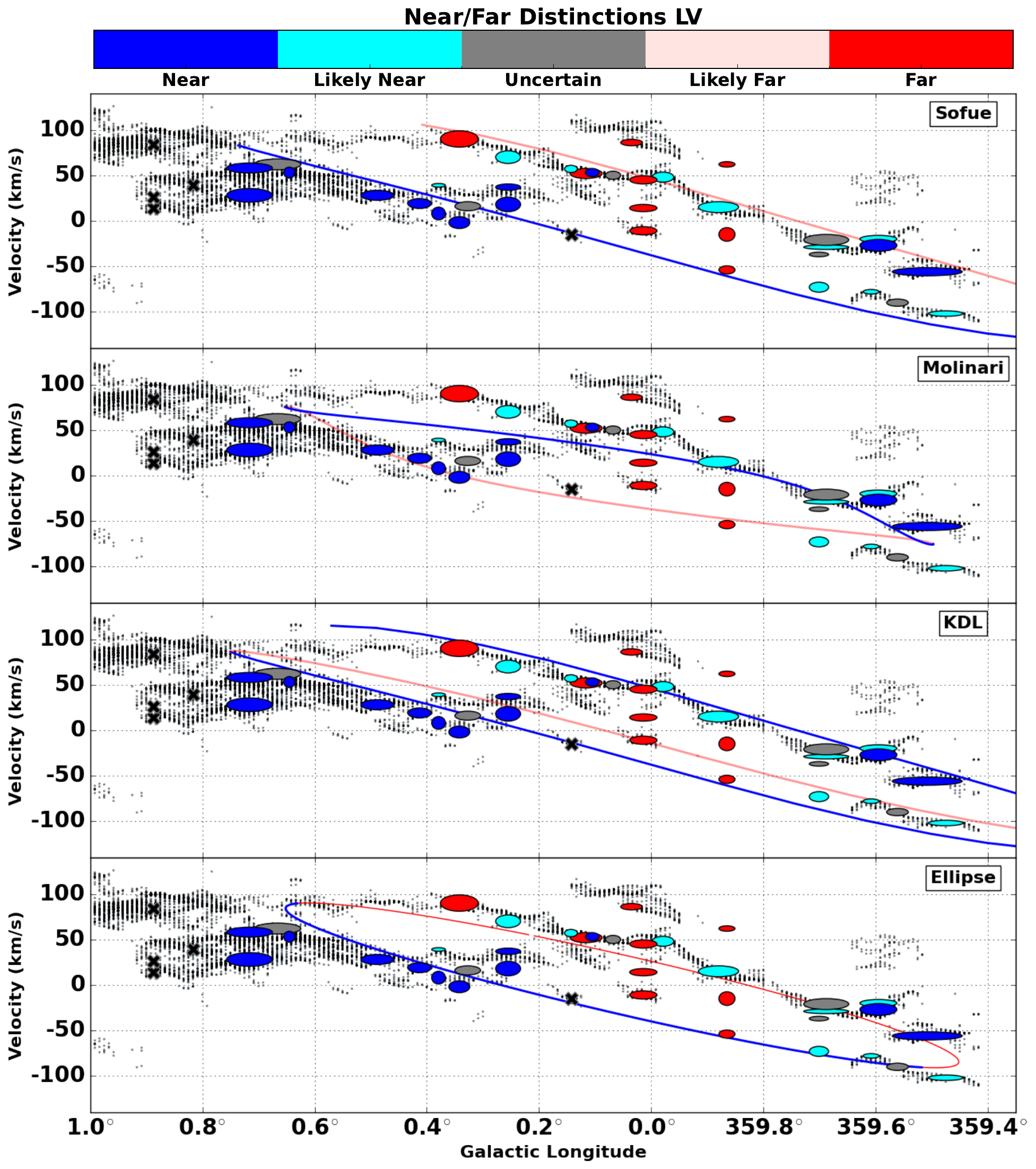}
    \caption{($\mathrm{\ell}$,v) projection of summarized near vs far likely positions of clouds in the CMZ ring based on comparison of the flux difference, flux ratio, 70\micron correlation coefficient, and absorption analysis. Catalog masks are plotted over the 2D projections of four orbital models in position-position space. From top to bottom: \citet{Sofue1995}, \citet{Molinari2011}, \citet{Kruijssen2015}, and a vertically oscillating elliptical orbit, similar to \citet{Molinari2011}. The colors correspond to Near (blue), Likely Near (cyan), Likely Far (rose), and Far (red) positions clouds based on visual summarizing of the four methods from Figure \ref{fig:compare_methods_3panel}. The horizontal extent of each point corresponds to the projected radius recorded in the dendrogram catalog table, while the vertical extent corresponds to the velocity dispersion. Clouds that show disagreement between two or more methods are marked Uncertain (gray). The grey background points correspond to the spectral decomposition of MOPRA HNCO data from \citet{Henshaw2016_GasKin_250pc}. Clouds for which near/far distinctions are not determined are denoted by gray crosses. The elliptical or KDL streams models show the best fits to the current summarized results, though it is not possible to confidently distinguish between them with the given data and methods.}
    \label{fig:LV_megaplot}
\end{figure*}

\begin{figure*}
    \centering
    \includegraphics[width=1\textwidth]{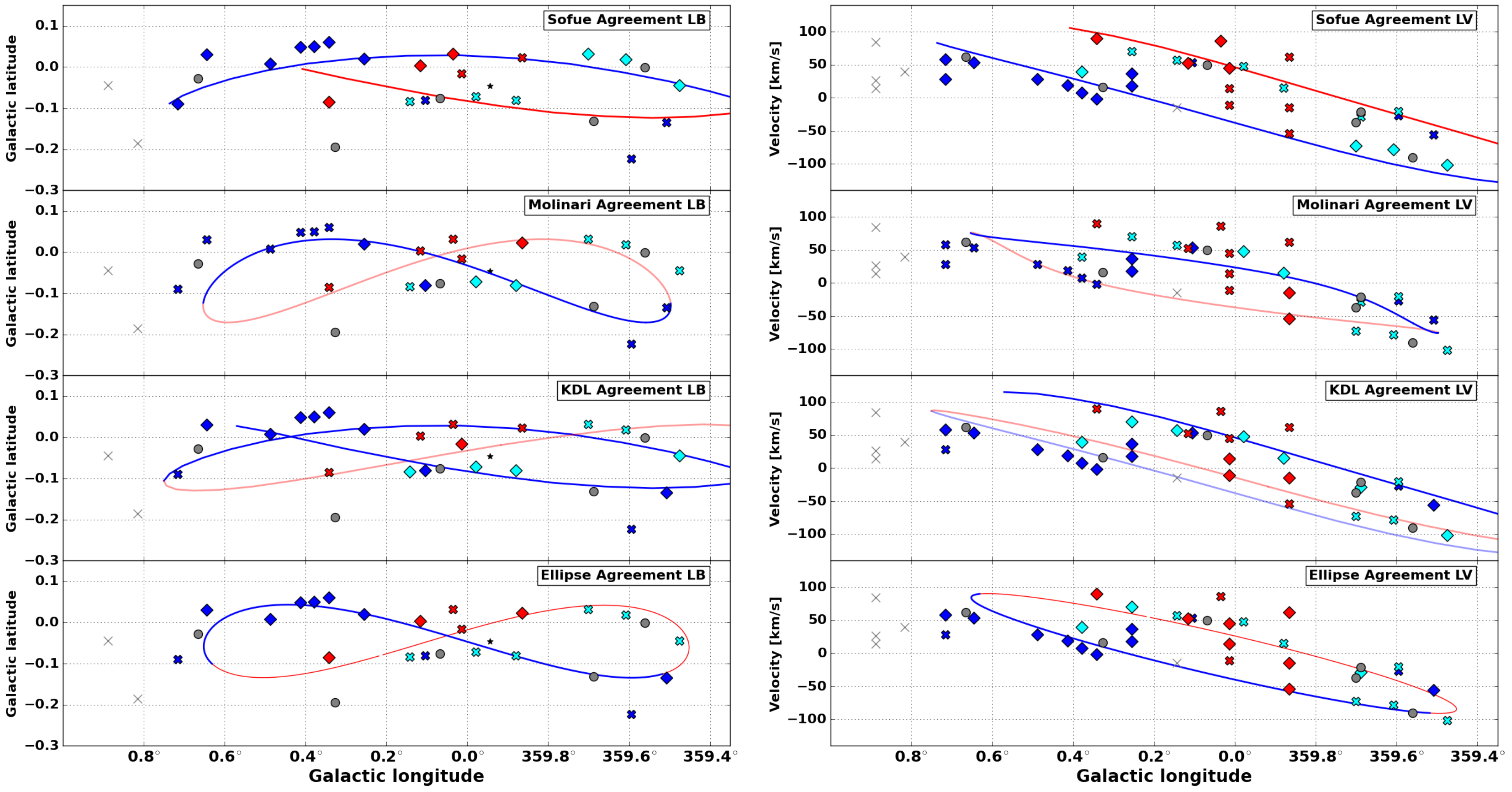}
    \caption{The minimum distance approach detailed in this paper favors the KDL model in terms of percent agreement between the orbital streams to the combined near/far distinctions from the three dust extinction methods and absorption method from \textit{Paper III}. This figure presents separate visualizations for the resulting fits in ($\ell$, $b$) (left) and ($\ell$, v) (right) space. Diamond markers note clouds which are deemed a "match" in both their $lbv$ positions and near/far distinctions. colored X's note clouds which do not match the streams.   Colors correspond to Near (blue), Likely Near (cyan), Likely Far (rose), and Far (red) positions of clouds based on visual summarizing of the four methods from Figure \ref{fig:compare_methods_3panel}. Clouds that show disagreement between two or more methods are marked Uncertain (gray circles). Clouds for which there are no near/far distinctions reported are marked by a gray X. The black star notes the location of SgrA*. The minimum distance approach ranks the model agreements as KDL (65\%), Ellipse (61\%), Sofue (55\%), Molinari (23\%).}
    \label{fig:percent_agree}
\end{figure*}

We report the results of our MIR dust extinction investigation as follows: 

\begin{itemize}

\item We find that high column densities and optical depths in the CMZ make traditional MIR dust extinction methods ineffective for determining line of sight positions of clouds in the CMZ. CMZ clouds appear almost entirely opaque at 8\micron, with a corresponding optical depth of $\tau_{8\mu\text{m} } = 5.47$, but are less optically thick in 70\micron with $\tau_{70\mu\text{m} } = 0.81$.


\item We introduce two new dust extinction methods to estimate the likelihood that a cloud is on the near or far side of the CMZ based on Spitzer 8\micron data. These methods use the flux difference and flux ratio of each cloud compared to a constant foreground and background.

\item We use a modified 70\micron extinction column density for a pixel-by-pixel comparison of calculated extinction column density to HiGAL emission for each clouds in the CMZ. We then compute a Pearson correlation coefficient to place clouds on the near side (r $\geq0.3$) or far side (r $< 0.3$) of the CMZ assuming a foreground fraction of 0.5 for all CMZ clouds. 

\item All three dust extinction methods share good agreement in near vs far distinctions, and also show agreement with absorption analysis discussed in \textit{Paper III} of this series.

\item We report near and far likelihoods for all molecular clouds in the CMZ based on our three different methods, presented in Table \ref{tab: flux_calc_tab}. 

\item Based on both qualitative and two quantitative summaries of the methods presented in this series, we compare our near/far position results with four CMZ geometric orbital models in both ($\mathrm{\ell}$,$b$) (Figure \ref{fig:LB_megaplot}) and ($\mathrm{\ell}$,v) (Figure \ref{fig:LV_megaplot}) projections.  Our conclusions for the projected fits of each model are as follows:

\textbf{Two spiral arms:} Our results are inconsistent with a projected two spiral arms orbital model based on \citet{Sofue1995}. In particular, the likely nearside positions of the 50 and 20 km/s clouds (IDs 10 and 9, respectively) are difficult to reconcile with the spiral arms model, though there also exists a possibility these clouds lie closer to SgrA*, rather than on the CMZ ring.

\textbf{Open stream:} Our near/far results for CMZ clouds are the most consistent with the open stream model presented in \citet{Kruijssen2015}. However, further work is needed to constrain the positions of clouds in the line of sight longitudes near SgrA*. Additionally, the KDL model still lacks an interpretation that explains its physical origin and connects the streams to the larger-scale flow, including the bar lanes that transport gas to the CMZ.

\textbf{Molinari et al.\ 2011 closed ellipse:} While the elliptical model proposed by \citet{Molinari2011} seems to match the near/far positions of clouds in ($\mathrm{\ell}$,$b$) space, this projected closed elliptical orbit shows significant inconsistencies in position-velocity space, particularly for the confidently near-sided dust ridge clouds. The kinematic inconsistencies strongly rule the Molinari ellipse out as a potential CMZ orbital model.

\textbf{Toy elliptical model:} \textit{Paper III} introduced a new elliptical orbit similar to \citet{Molinari2011}, which assumes a constant $z$-component of the angular momentum and is offset from SgrA* at ($\ell$,$b$) = (0.05\deg, -0.0462\deg). While the Molinari ellipse may not fit the data well in ($\mathrm{\ell}$,v) space, the modified toy-model ellipse shows much better agreement with the near/far distinctions presented in this series. Our results suggest that a modified closed elliptical orbit may still be valid. However, development of a more physical elliptical model is still needed.

\item We find that all current CMZ orbital models lack the complexity needed to describe the motion of gas in the CMZ. Further work is needed to produce a more complex model to accurately describe the gas flows and their connection to large-scale dynamics. The use of proper motion measurements, x-ray echoes, and potential line-of-sight distance measurements to individual clouds from NIR dust extinction techniques using JWST data can help constrain the likely near/far positions of clouds, and the creation of a more complex, physically informed model of the gas flows in the CMZ.

\end{itemize}

\section*{Acknowledgements}

D.\ Lipman gratefully acknowledges funding from the National Science Foundation under Award Nos. 1816715, 2108938, and CAREER 2145689, as well as from the NASA Connecticut Space Grant Consortium under PTE Federal Award No: 80NSSC20M0129.

C.\ Battersby  gratefully  acknowledges  funding  from  National  Science  Foundation(NSF)  under  Award  Nos. 1816715, 2108938, 2206510, and CAREER 2145689, as well as from the National Aeronautics and Space Administration through the Astrophysics Data Analysis Program under Award ``3-D MC: Mapping Circumnuclear Molecular Clouds from X-ray to Radio,” Grant No. 80NSSC22K1125. 

AG acknowledges support from the NSF under grants AST 2008101, 2206511, and CAREER 2142300.

JDH gratefully acknowledges financial support from the Royal Society (University Research Fellowship; URF/R1/221620). 

Herschel is an ESA space observatory with science instruments provided by European-led Principal Investigator consortia and with important participation from NASA.

M. C. Sormani acknowledges financial support from the European Research Council under the ERC Starting Grant ``GalFlow'' (grant 101116226).

SCOG and RSK acknowledge funding from the European Research Council via the ERC Synergy Grant ECOGAL (grant 855130), from the Heidelberg Cluster of Excellence STRUCTURES in the framework of Germany's Excellence Strategy (EXC-2181/1 - 390900948), and from the German Ministry for Economic Affairs and Climate Action in project MAINN (funding ID 50OO2206).

E.A.C.\ Mills  gratefully  acknowledges  funding  from the National  Science  Foundation  under  Award  Nos. 2206509 and CAREER 2339670.

Q. Zhang acknowledges the support from the National Science Foundation under Award No. 2206512.

COOL Research DAO is a Decentralized Autonomous Organization supporting research in astrophysics aimed at uncovering our cosmic origins.

\software{Astropy \citep{astropy:2013, astropy:2018, astropy:2022}}, Scipy \citep{Scipy2020}.
This research made use of SAO Image ds9 \citep{Joye2003}.
This research also made use of Montage. It is funded by the National Science Foundation under Grant Number ACI-1440620, and was previously funded by the National Aeronautics and Space Administration's Earth Science Technology Office, Computation Technologies Project, under Cooperative Agreement Number NCC5-626 between NASA and the California Institute of Technology.

\clearpage

\appendix 
\counterwithin{figure}{section}

The following Appendices cover our investigation of applying typical dust extinction methods to the high column density regions in the CMZ. 

Section \ref{Appendix_typical_NIR_MIR} discusses typical NIR and MIR extinction mapping techniques used to determine distances in the Galactic Disk. In Section \ref{Appendix_MIR_limitations} we explore the application of 8\micron extinction column density calculations to CMZ clouds, and show that these techniques are not appropriate for use in the CMZ. 

In Section \ref{Appendix_8vs70} we compare the relative limitations of the 8\micron extinction calculation to the modified 70\micron extinction calculation utilized in this paper, and conclude that the 70\micron method is a clear improvement and is appropriate to use as a technique to determine near/far positions of molecualr clouds in the CMZ.

Section \ref{Appendix_all_cloud_corr_coeffs} reports 70\micron calculated extinction vs observed emission column density plots for all CMZ clouds, as discussed in Section \ref{sec:70um_ext_method}.

\section{Typical NIR and MIR Extinction Mapping Methods in the Galactic Disk} \label{Appendix_typical_NIR_MIR}

Various dust extinction methods have been used to create detailed maps of nearby star-forming regions in the Milky Way \citep{Zhang2022, Kauffmann2008, Flaherty2007, Lombardi2014} and external galaxies \citep{Kainulainen2007, Faustino_Vieira2023}. In particular, NIR dust extinction has been used to probe the 3-D structure of nearby star-forming regions \citep[e.g.][]{Rezaei2022}. NIR color excess methods often make assumptions of universal NIR extinction laws, which tend to yield higher uncertainties in denser regions \citep{Kainulainen2007, Lombardi2014}. Photometric color-excess methods have also been combined with parallax-based techniques using Gaia measurements of embedded or background stars to create 3-D maps of nearby interstellar dust \citep{Lallement2019, Chen2019, Rezaei2022}. However, as discussed in Section~\ref{sec:intro}, the gas surface densities in the CMZ are too high for NIR dust extinction methods alone. In fact, \citet{Kainulainen(2013)} developed a combination of techniques using NIR photometry on the edges of clouds and MIR extinction in the more optically thick centers, which results in a temperature-independent method for mapping nearby IRDCs. Since the densities in the CMZ tend to be far greater than those of nearby clouds, it is likely MIR techniques will be more applicable to the CMZ. Thus, in this paper, we choose to investigate the use of MIR dust extinction techniques similar to \citet{ Battersby2010} and \citet{Ellsworth-Bowers2013}, using Spitzer 8\micron data. 


As previously mentioned in Section~\ref{sec: dustext_optical_depth}, Spitzer 8\micron images have been used to obtain extinction column density maps for IRDCs in the disk \citep{Battersby2010, Ragan2009, Butler&Tan2009} and to  determine probabilistic near vs.\ far kinematic distances of sources using 8\micron extinction features in the Galactic plane \citep{Ellsworth-Bowers2013}. These methods assume the extinction and emission column densities should correlate well, and hence this technique works best for non-opaque sources. However, clouds that appear optically thick in the MIR should still show an easily detectable difference between the MIR absorption features and the overall foreground and background intensity in the absence of absorption, even when assuming most of the emission is due to the foreground \citep{Ellsworth-Bowers2013}. 

\section{8\micron Dust Extinction Limitations in the CMZ}\label{Appendix_MIR_limitations}

As discussed in the text, the extreme brightness of the Galactic Center in 8\micron should make it possible to differentiate clouds in front of or behind the bright emission. However, high column densities, as well as complicated emission features, in the CMZ create potential difficulties when attempting to apply dust extinction column density methods to clouds in the CMZ.
In this Section we investigate the limitations of the 8\micron dust extinction column density method when applied to the CMZ. 

Typical dust extinction methods applied to molecular clouds in the disk of the Galaxy aim to compare emission column density with calculated extinction column densities to determine a ratio between emission and extinction in the infrared \citep{Battersby2010, Butler&Tan2009}. For example, \citet{Battersby2010} compare Bolocam 1.1 mm dust emission and Spitzer 8\micron data for IRDCs in the disk of the Galaxy. We explore the same methodology to calculate extinction-based surface densities of clouds at 8\micron using Equation 7 from \citet{Battersby2010}:

\begin{equation} \label{eqn:ExtSig}
    \Sigma = -\frac{1}{\kappa_{\nu}} \ln \left[\frac{(s+1)\mathrm{I}_{\nu 1, \rm obs} - (s+1)f_{\rm fore}\mathrm{I}_{\nu 0,\rm obs}}{(1-f_{\rm fore})\mathrm{I}_{\nu 0,\rm obs}}\right],
\end{equation}
where $\kappa_{\nu}$ is the dust opacity, which at $8\mu m$ is taken to be 11.7 cm$^{2}$g$^{-1}$, as used in \citet{Battersby2010} based on the models for \citet{Ossenkopf&Henning(2004)} of thin ice mantles that have undergone coagulation for $10^{5}$ years at a density of $\sim 10^{6}$cm$^{-3}$. The scattering coefficient, s, is a correction for the Spitzer IRAC array, which systematically increases the surface brightness of an extended source by $\sim 30\%$\footnote{\protect\url{https://irsa.ipac.caltech.edu/data/SPITZER/docs/irac/}}. Thus we assume a scattering coefficient of $\mathrm{s} = 0.3$. The resulting mass surface density, $\Sigma$, is dependent on the measured intensity in front of the cloud (I$_{\nu 1,\rm obs}$), the composite intensity of the foreground and estimated background at the location of the cloud (I$_{\nu 0,\rm obs}$), and the foreground intensity ratio (f$_{\rm fore}$), defined as the ratio of intensities out to 8 kpc and out to 16 kpc (i.e.\ twice the distance to the CMZ, assuming a distance to SgrA* of $\sim$ 8 kpc).  

 We assume the distribution of dust emission varies symmetrically with line of sight distance to the CMZ. Both \citet{Butler&Tan2009} and \citet{Battersby2010}  estimate f$_{\rm fore}$ by assuming the Galactic distribution of hot dust is the same as the distribution of the Galactic surface density of OB associations \citep{McKee&Williams(1997)}:
\begin{equation}\label{eqn:OB_assoc}
    \Sigma_{\rm OB} \propto \text{exp} \left( - \frac{R}{H_{R}} \right)
\end{equation}
where R is the galactocentric radius and H$_{R}$ is the radial scale length. However, \citet{Butler&Tan2009} note that while the foreground fraction is indeed necessary, estimations of f$_{\rm fore}$ hold a large source of uncertainty due to small scale spatial variations in the hot dust emission in the Galaxy, and thus limit their extinction mapping analysis to relatively nearby clouds. 

We implement this method and test its viability in the CMZ by assuming that half of the observed intensity is due to foreground emission, i.e.\ $f_{\rm fore} =0.5$. We make this assumption as a test, as all catalogd clouds exist within a central few hundred parsec radius of the Galactic Center, meaning differences in f$_{\rm fore}$ between clouds should be small, and likely within a range from $0.45 <$ $f_{\rm fore} < 0.55$ for most CMZ clouds. 

To measure I$_{\nu 0,\rm obs}$, we use the smoothed CMZ model discussed in Section \ref{sec:flux_methods}, and follow the same general procedure at Section \ref{sec:70um_ext_method} applied to the 8\micron map. The result of these calculations produces an extinction column density map for each cloud mask in the catalog. The $N(\rm H_{2})_{8\mu m}$ map is then smoothed to match the 36\arcsec FWHM of the HiGAL column density map in order to produce an ``observed" comparison to the HiGAL column density maps for each cloud.

\begin{figure}
    \includegraphics[width=0.5\textwidth]{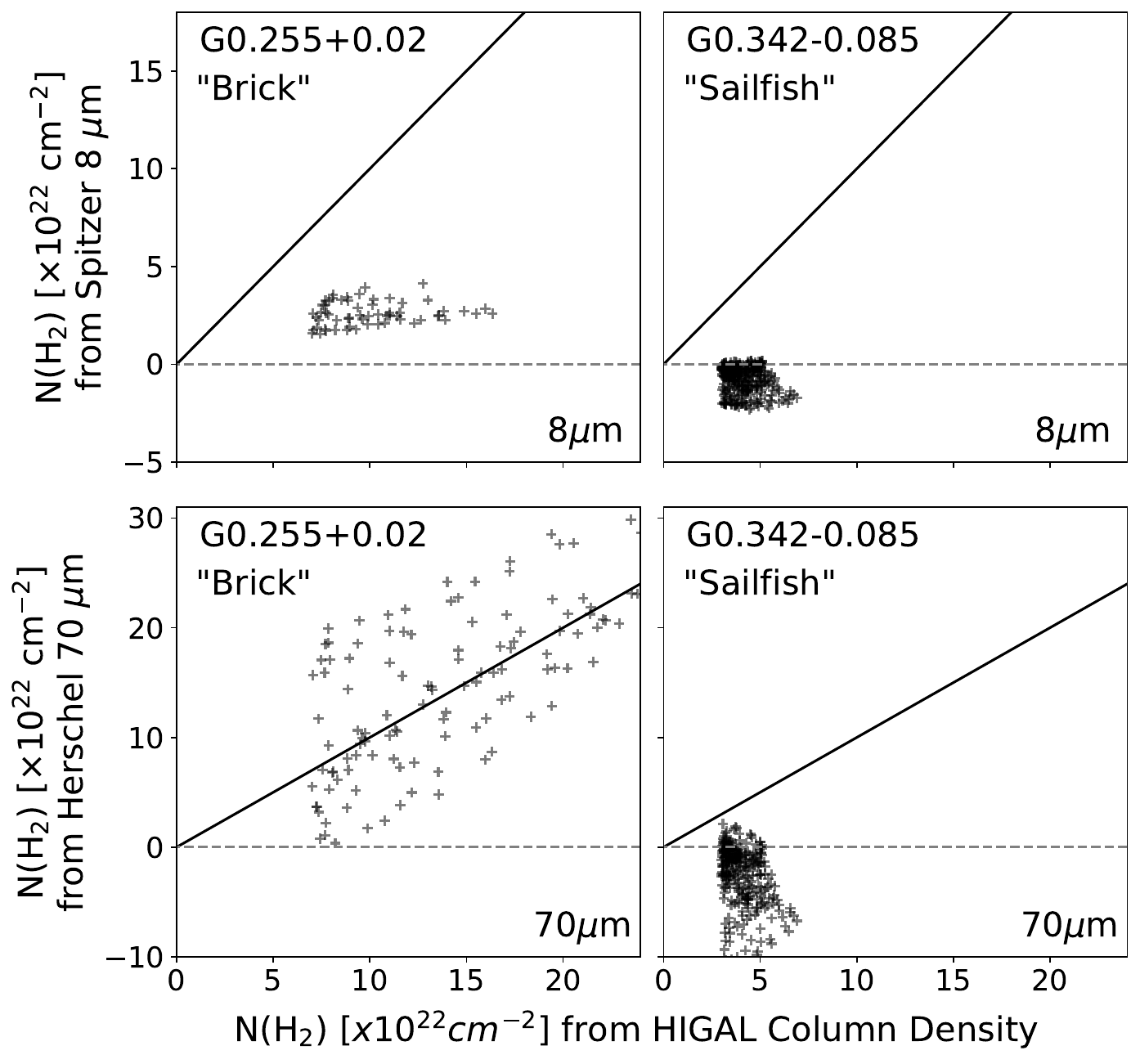}
    \caption{High column densities in the CMZ mean clouds become opaque in the Spitzer 8\micron, leading to almost immediate saturation in the extinction column density; Herschel 70\micron data appears less opaque and shows good correlation with integrated column densities. Top row: example of 8\micron calculated extinction column density and emission from HiGAL for the Brick (left) and Sailfish (right) using $f_{\rm fore}=0.5$. The solid black line shows a 1:1 trend. Far-side clouds such as the Sailfish are expected to show no correlation between the extinction and emission, as seen here. Clouds thought to be in front of the bright galactic emission, such as the Brick, should show some correlation. Bottom row: The same example clouds, but using 70\micron calculated extinction column density. }
    \label{fig:70vs8_brick_sailfish_comparison_4panel}
\end{figure}

The top panels of Figure \ref{fig:70vs8_brick_sailfish_comparison_4panel} show an example comparison of the column densities inferred from 8\micron extinction and HiGAL emission for the Brick (left) and G0.342-0.085 (hereafter, the Sailfish) (right) using $f_{\rm fore}=0.5$ (i.e.\  half of the observed emission from the composite foreground and background at the location of the cloud is assumed to be from the foreground). The Sailfish still shows a negative correlation between the 8\micron extinction and emission, indicating the densest regions of the cloud are unlikely to be extincting light from behind. However, the Brick only shows a weak correlation, and subsequently highlights an important limitation typical MIR extinction method faces when applied outside of nearby regions in the disk: extreme densities and opacity.

Column densities in the CMZ are of order $\sim 10^{23}$~cm$^{-2}$, corresponding to A$_{V}$ magnitudes $\sim 50$~mag. This means molecular clouds in the CMZ region appear almost entirely opaque at 8\micron, with a corresponding optical depth of $\tau_{8\mu\text{m}  } \sim 5$, leading the extinction column densities to immediately level off with a relatively small range, while the emission column densities continue to grow. The strong saturation occurs for the majority of CMZ clouds in 8\micron and makes it difficult to determine significant extinction-emission correlations for any f$_{\rm fore}$ within the range of the CMZ. Additionally, densities in the CMZ are much higher than the disc; the fitted HiGAL column densities tend to sit almost twice as high as the calculated extinction column density.

The optical depth of clouds also relates to another limitation of the extinction mapping method in the CMZ. The assumption of \ffore = 0.5 means that the light reaching us from the direction of the cloud can be no less than half of the background light. Clouds like the Brick extinguish significantly more than 50\% of the background light, breaking the model and leading to a negative value inside of the logarithmic term. The large opacity means that essentially all light from behind the cloud is extinguished, so the choice of \ffore, with its attendant uncertainties, ends up determining the cloud extinctions. This issue is less prominent in less extinguished regions where imperfections in \ffore\xspace and the background lead to more modest errors in the derived extinctions. A visual example of this limitation for the Brick cloud can be seen in Figure \ref{fig:6_panel_brick_8um}. Panels (a) and (c) show the Spitzer 8\micron and smoothed 8\micron maps of the Brick, respectively. Most of the darkest areas of the cloud have observed intensities that are less than half of the modeled background, leading to most of the cloud having nonsensical extinction values in panel (d). While the Brick is the most extreme example of this issue for CMZ clouds, it highlights how Equation \ref{eqn:ExtSig} cannot be used for opaque areas that would be of most interest in the extinction-emission column density comparisons.

Another important limitation of the calculated 8\micron extinction method can be attributed to the Spitzer 8\micron filter. Equation \ref{eqn:ExtSig} assumes the cloud itself is not a source of 8\micron emission. However, the Spitzer 8\micron filter is broad enough to include the 7.7\micron PAH emission feature. For IRDCs in the disk, where the interstellar radiation field is lower and non-variable over the surface of the clouds, the strength of this feature is small and so it is not a significant issue for this approximation in areas that are not close to active star formation regions. However, in the CMZ, where the ISRF is known to be much higher compared to the disk, the existence of strong PAH features in 8\micron filter cannot be ignored, 
and our assumption that the cloud itself is not an 8\micron source can break down. This
likely contributes to issues in the applicability of the calculated 8\micron extinction method in the CMZ.

\begin{figure*}[!htb]
    \centering
    \includegraphics[width=1\textwidth]{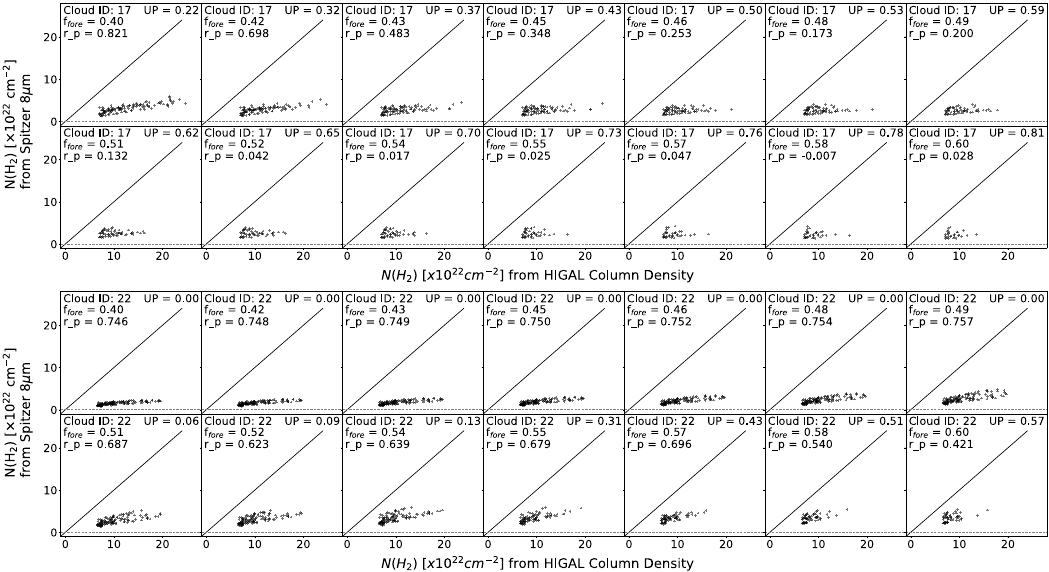}
    \caption{Comparison of 8\micron calculated extinction column density and emission from HiGAL for various foreground fractions between 0.4 and 0.6 for ID 17 (the Brick; top) and ID 22 (bottom). Annotated values include the Pearson correlation coefficient between the extinction and emission (r), as well as the fraction of undefined pixels (UP) in the extinction calculation due to limitations of Equation \ref{eqn:ExtSig} in high column density areas. As with most CMZ clouds, the column densities in the Brick are high enough that the cloud saturates quickly in 8\micron, regardless of the foreground fraction, making it difficult to use the 8\micron method to distinguish best-fit distances.}
    \label{fig:IDs17and22_8um_ExtN_ffore_subset}
\end{figure*}

The optical depth and 8\micron filter features mean the MIR extinction technique cannot be used anywhere within the range of the CMZ to infer line of sight locations. Varying f$_{\rm fore}$ should result in visibly different correlations between the calculated extinction column densities and the observed emission column densities. However, as discussed in the text, CMZ clouds become saturated in 8\micron, regardless of the value for f$_{\rm fore}$. Figure \ref{fig:IDs17and22_8um_ExtN_ffore_subset}  show a comparison of 8\micron calculated extinction column density and emission from HiGAL for various foreground fractions between 0.4 and 0.6 for the dendrogram structures with ID 17 (the Brick; top) and ID 22 (bottom). Additionally, the fraction of undefined pixels (UP) is also noted for each f$_{\rm fore}$. The undefined pixels occur due to negative values inside of the logarithmic term of Equation \ref{eqn:ExtSig}, which occur if the measured intensity of the pixel is less than the estimated contribution from the foreground. As seen in the examples from Figure \ref{fig:IDs17and22_8um_ExtN_ffore_subset}, the extinction caps off quite quickly, when we would expect the brightest emission points to scale better with calculated extinction. Confidently distinguishing between different values of \ffore\xspace becomes difficult—especially as more data points are lost to UP in the densest and brightest parts of the clouds due to the limitations of the equation. The undefined pixels are often the densest regions of the clouds that should return the highest extinction column densities. 

Large numbers of undefined pixels are indicative of an incorrect estimate for the foreground contribution. One way this could happen is if the value adopted for f$_{\rm fore}$ is too large. If we assume that the distribution of foreground emission simply tracks the distribution of OB associations in the Galactic disk (Equation \ref{eqn:OB_assoc}), then there is a direct mapping between the value of f$_{\rm fore}$ and the distance to the cloud. For example, if we were to adopt a value of $f_{\rm fore} = 0.6$ for the Brick and Cloud 22, it would place both of them $> 800$~pc behind Sgr~A*. However, as we see in Figure \ref{fig:IDs17and22_8um_ExtN_ffore_subset}, this results in a large fraction of unphysical column densities (i.e.\  undefined pixels) for both clouds, suggesting that they are not at such a distance. Reducing f$_{\rm fore}$ reduces the number of undefined pixels, at the cost of moving the clouds closer to us along the line of sight. For example, if we adopt $f_{\rm fore} = 0.4$, this corresponds to a distance of $\sim 540$~pc in front of Sgr~A*. At this distance, Cloud 22 no longer has any undefined pixels, but around 22\% of the pixels in the Brick remain undefined. The Brick has been confidently constrained to lie on the near side of the Galactic Center, but still within the CMZ \citep{Nogueras2021}, making a value of $f_{\rm fore} < 0.4$ for the Brick highly improbable. Therefore, an incorrect choice for f$_{\rm fore}$ cannot be the only reason for our estimate of the foreground being incorrect. This results in the 8\micron method being either 1.) unable to reliably calculate the extinction column density and determine a corresponding f$_{\rm fore}$ in opaque regions, or 2.) heavily impacted by the loss of densest and brightest areas of a cloud (often towards the centers), which would be important areas of interest for the extinction correlations. We do note that it is maybe possible to use \ffore in a similar way as the flux ratio method to approximate near/far positions instead of provide confident line of sight distance estimates. However, we choose to explore modifying the 8\micron method for less optically thick wavelengths, and leave development of an \ffore~based near/far distinction approach for a future paper.

\subsection{Comparison of 8\texorpdfstring{$\mu$}{u}m\xspace and application 70\texorpdfstring{$\mu$}{u}m\xspace calculated extinction in the CMZ}\label{Appendix_8vs70}

Both the saturation and undefined pixels make it difficult to use the 8\micron method as a way to determine a most probable distance to the cloud based on either a correlation coefficient or fit to the 1:1 line. Thus, we introduced the modified 70\micron method in this work.

The bottom panels of Figure \ref{fig:70vs8_brick_sailfish_comparison_4panel} show the results of the 70\micron method applied to the Brick (left) and Sailfish (right). The Sailfish cloud, as expected, still shows no correlation between the extinction and emission. However, the Brick shows a visibly reasonable correlation between the calculated extinction and HiGAL emission. 

\begin{figure*}
    \includegraphics[width=\textwidth]{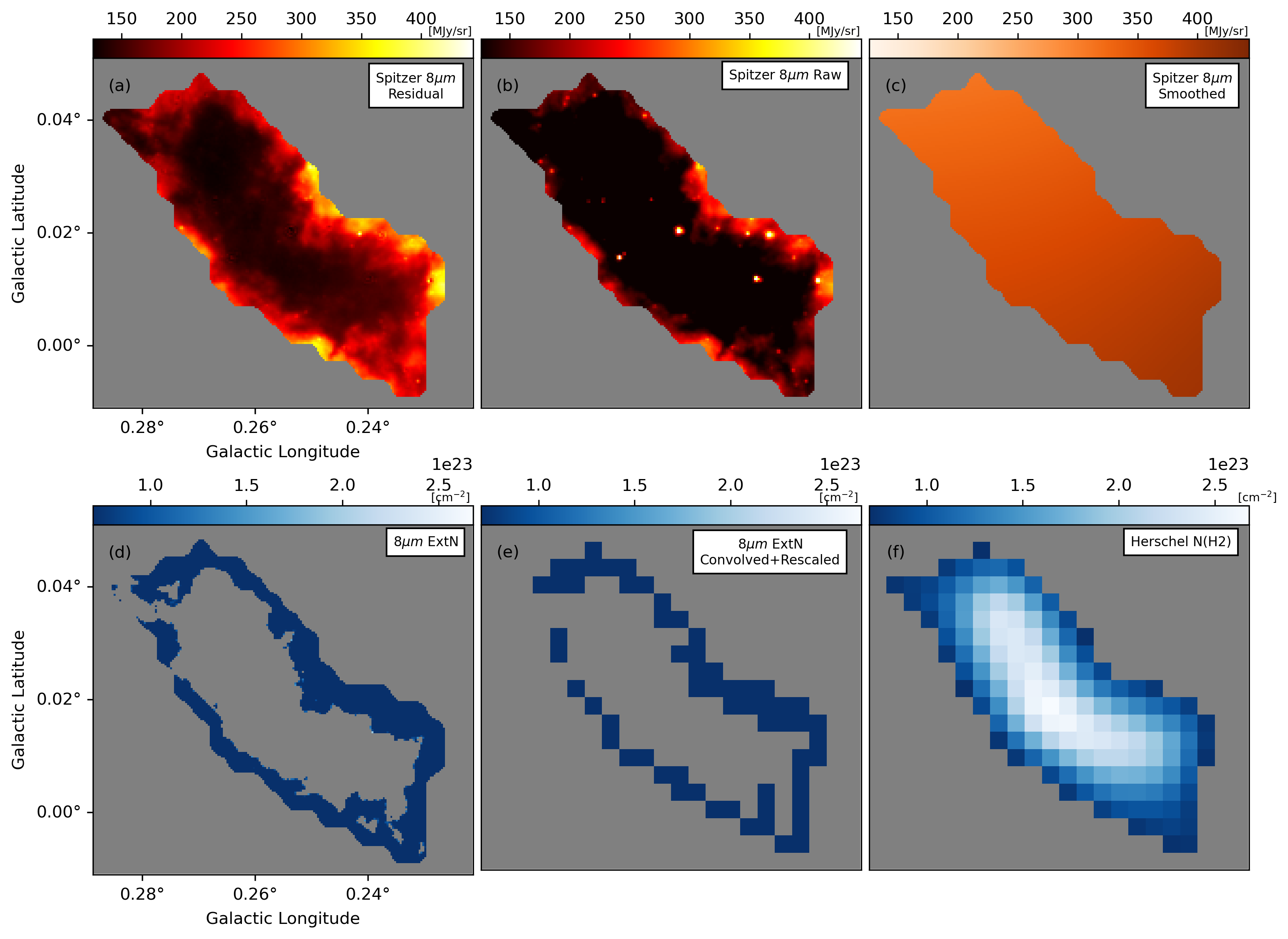}
    \caption{The high column density of clouds at 8\micron causes typical extinction methods to fail in opaque areas of clouds where the modeled background is much higher than observed intensity. The extinction cannot be calculated for these pixels, as it is only possible to place a lower limit on $\tau$. This figure show examples of the 8\micron extinction method for the Brick. Each panel shows a cutout mask of the cloud applied to the full maps. The top panels show (from left to right) (a) Spitzer 8\micron residual map, (b) raw Spitzer 8\micron (no subtraction of foreground stars), (c) the smoothed 8\micron. The bottom panels (from left to right) show (d) calculated 8\micron extinction column density with $f_{\rm fore} = 0.5$, (e) the same map from panel (d) convolved to the HiGAL column density beam and regrided to match the HiGAL pixel scaling, (f) HiGAL column density. Gray pixels are NaN values.}
    \label{fig:6_panel_brick_8um}
\end{figure*}

\begin{figure*}
    \includegraphics[width=\textwidth]{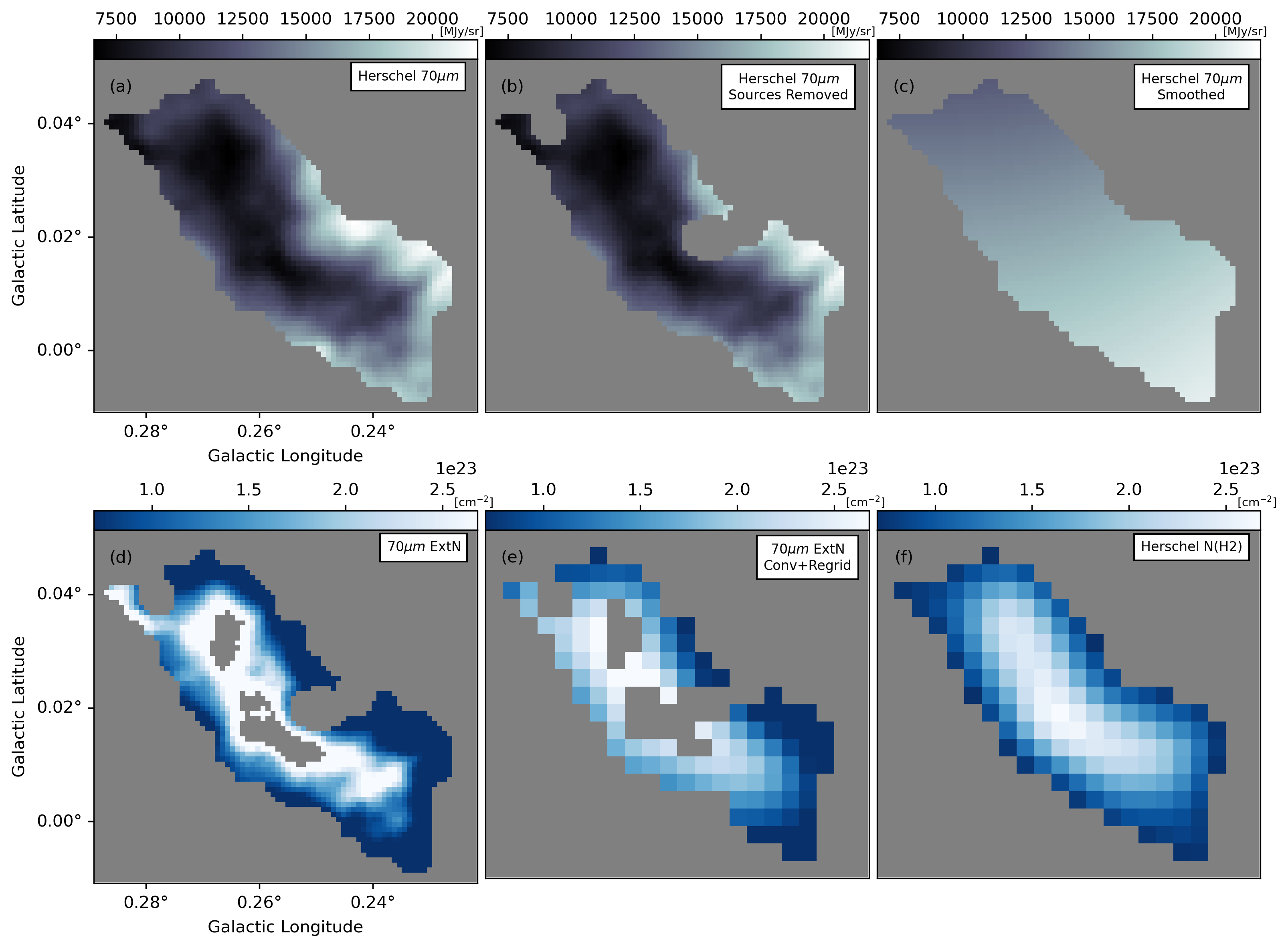}
    \caption{Our new 70\micron calculated extinction column density technique shows significantly better results compared to the typical 8\micron method, due clouds being optically thin in 70\micron. Each panel shows a cutout mask of the cloud applied to the full maps. The top panels show (from left to right) (a) Herschel 70\micron map, (b) the Herschel 70\micron map with catalogd 70\micron sources removed, (c) the smoothed 70\micron. The bottom panels (from left to right) show (d) calculated 70\micron extinction column density with $f_{\rm fore} = 0.5$, (e) the same map from panel (d) convolved to the integrated HiGAL column density beam and regridded to match the map's pixel scaling, (f) HiGAL column density. Gray pixels are NaN values.}
    \label{fig:6_panel_brick_70um}
\end{figure*}

Additionally, we visually compared the optical depth limitations in the densest regions of the Brick for both the calculated 8\micron and 70\micron extinction methods. Figures \ref{fig:6_panel_brick_8um} and \ref{fig:6_panel_brick_70um} show a visual representation of the typical 8\micron method and our new 70\micron method applied to the Brick for $f_{\rm fore} = 0.5$. Panel (d) of Figure \ref{fig:6_panel_brick_8um} shows the calculated 8\micron extinction column density for the cloud, with an extreme number of undefined pixels in the center where the cloud is most optically thick. The number of undefined pixels makes a comparison of the calculated extinction with the emission from Herschel nearly impossible for the cloud. On the other hand, Panel (d) of Figure \ref{fig:6_panel_brick_70um} shows the 70\micron calculated extinction column density for $f_{\rm fore} = 0.5$ for the Brick. There are still areas of undefined pixels in the highest opacity regions. An $f_{\rm fore} < 0.45$ would be needed to completely avoid any undefined pixels that are not due to the removal of 70\micron sources. This $f_{\rm fore}$ value would place the Brick  $\sim 180$~pc in front of SgrA*, which would comfortably place the Brick within the CMZ. 

The 70\micron method shows significant improvement compared to the 8\micron method, and it is still possible to compare the extinction and Herschel emission despite NaN pixels from both the source removal and undefined pixels. Additionally, the Brick is one of the most extreme cases of these limitations in the CMZ. Other clouds show fewer opacity issues, as discussed in the main text. Overall, this demonstrates that our proposed 70\micron method is a better choice for dust extinction mapping for clouds in the high density CMZ.

\clearpage 

\section{70\texorpdfstring{$\mu$}{u}m\xspace Extinction vs Emission Correlation Coefficients for all CMZ clouds}\label{Appendix_all_cloud_corr_coeffs}

\begin{figure*}
    \centering    \includegraphics[width=0.92\textwidth,height=\textheight,keepaspectratio]{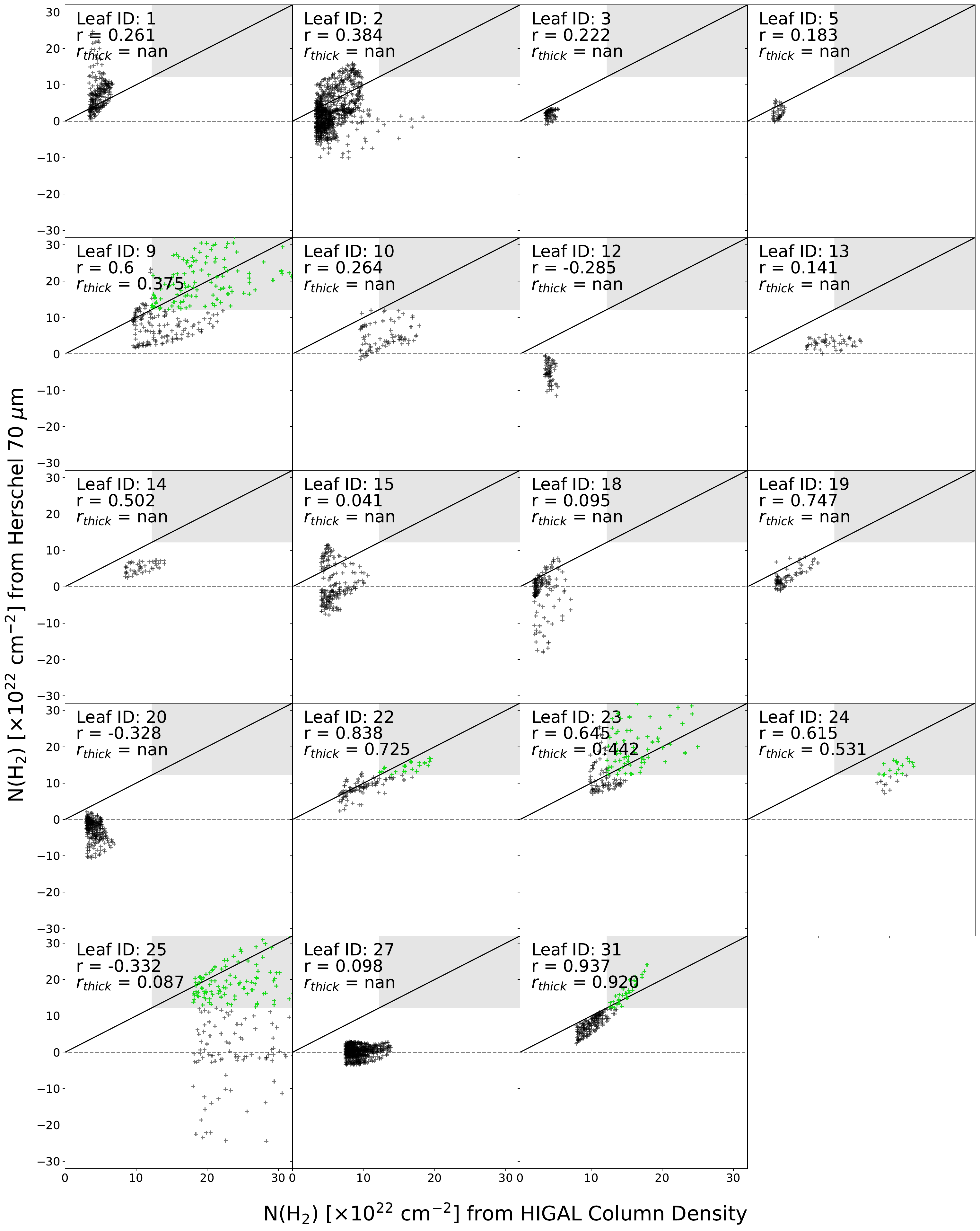}
    \caption{Herschel 70\micron calculated extinction column density compared to the emission column density from HiGAL using $f_{\rm fore}=0.5$ for catalog clouds with single velocity components. The solid black line shows a 1:1 trend. A Pearson correlation coefficient of r $\geq0.3$ defines a reasonable correlation between extinction and emission column densities. We determine that reasonably correlated clouds are likely on the near side, in front of the bright galactic emission. Likewise, uncorrelated clouds with r $< 0.3$ are deemed to be on the far side of the CMZ. The gray shaded area indicates the optically thick region for 70\micron. Green crosses within the gray shaded region correspond to points which are past the 70\micron optically thick limit on both axes, for which a thick correlation coefficient $\mathrm{r}_{thick}$ is reported as well.} 
    \label{fig:corr_coeff_single_comp}
\end{figure*}

The 70\micron extinction column density described in Section \ref{sec:70um_ext_method} was calculated for all clouds in the CMZ. We use $f_{\rm fore}=0.5$ for all clouds in the catalog. Figure \ref{fig:corr_coeff_single_comp} shows the Herschel 70\micron calculated extinction column density compared to the HiGAL emission column density for 19 out of 31 clouds which reported a single velocity component from HNCO spectral fittings. 

\begin{figure*}[htb!]
    \centering
    \includegraphics[width=\textwidth,height=\textheight,keepaspectratio]{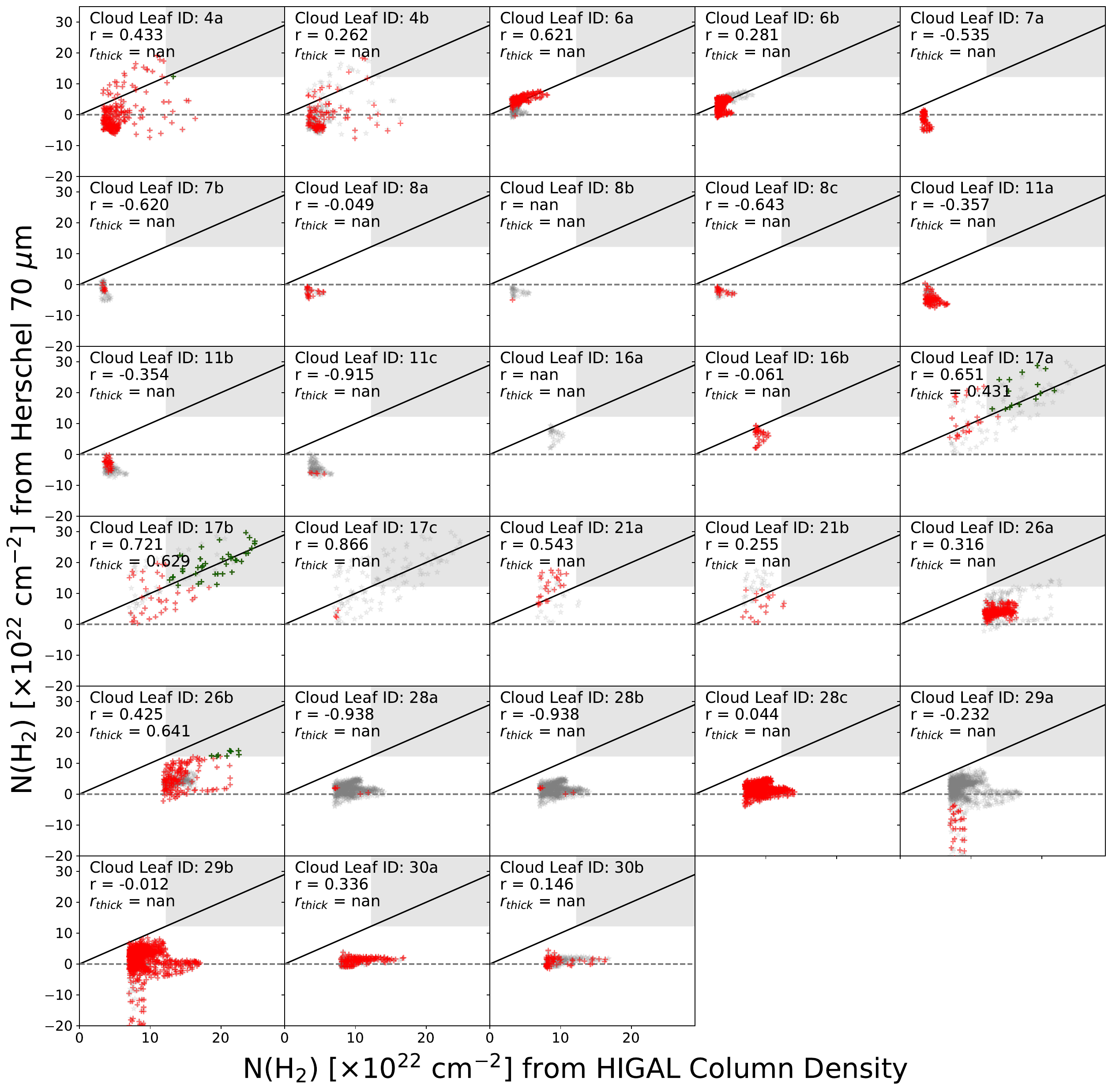}
    \caption{Herschel 70\micron calculated extinction column density compared to the emission column density from HiGAL using $f_{fore}=0.5$ for catalog clouds with multiple velocity components. Points belonging to the overall cloud mask are plotted in gray. The overplotted red points indicate the pixels that belong to the individual submask ID in each panel. The solid black line shows a 1:1 trend. A Pearson correlation coefficient of r $\geq0.3$ defines a reasonable correlation between extinction and emission column densities. We determine that reasonably correlated clouds are likely on the near side, in front of the bright galactic emission. Likewise, uncorrelated clouds with r $< 0.3$ are deemed to be on the far side of the CMZ. The gray shaded area indicates the optically thick region for 70\micron. Green crosses within the gray shaded region correspond to points which are past the 70\micron optically thick limit on both axes, for which a thick correlation coefficient $\mathrm{r}_{thick}$ is reported as well. ID 16a has an empty peak intensity mask, and therefore has no calculated r.} 
    \label{fig:corr_coeff_submasks}
\end{figure*}

A comparison of the extinction and emission column densities for clouds with multiple velocity components is shown in  Figure \ref{fig:corr_coeff_submasks}. As noted in Section \ref{sec:data}, the submasks were generated by creating a peak intensity map of the HNCO emission within a velocity range defined by the velocity dispersion of the component about the centroid velocity. Black points show the pixel-by-pixel extinction vs emission for the full leaf mask (similar to Figures \ref{fig:corr_coeff_single_comp} and \ref{fig:corr_coeff_brick_fish_50}. The red points indicate the pixels that belong to the individual submask ID in each panel. The reported r  is used to determine if each submask should be placed in a likely near or likely far distinction. ID 16a has an empty peak intensity mask (i.e. there were no pixels assigned to the mask based on the velocity range), and therefore has no calculated r. Differing near/far distinction between sub-masks provides strong indication that certain catalog leafs have separate substructure based on their velocities, such as with IDs 6a and 6b; IDs  26a and 26b; and IDs 17a, 17b, and 17c. 

\clearpage

\bibliography{references.bib}{}
\bibliographystyle{aasjournal}

\end{document}